\documentclass[aps,prd,showpacs,nofootinbib,floats,floatfix,reprint,preprintnumbers,groupedaddress,twocolumn]{revtex4}

\usepackage[normalem]{ulem} 
\usepackage{bm}
\usepackage[titletoc]{appendix}
\usepackage{latexsym}
\usepackage{dcolumn}
\usepackage{amsmath,amsfonts,amssymb}
\usepackage{graphicx,epsfig}
\usepackage{amsmath}
\usepackage{fancyhdr}
\usepackage{hyperref}
\usepackage{graphicx,epstopdf}
\usepackage{xcolor}
\hypersetup{
	colorlinks   = true, 
	urlcolor     = blue, 
	linkcolor    = blue, 
	citecolor   = red 
}

\def\bea{\begin{eqnarray}}
\def\eea{\end{eqnarray}}

\begin{document}
\title{Study of relativistic accretion flow around KTN black hole with shocks}
\author{Gargi Sen}\email{g.sen@iitg.ac.in}
\author{Debaprasad Maity}\email{debu@iitg.ernet.in}
\author{Santabrata Das}\email{sbdas@iitg.ac.in (Corresponding Author)}
\affiliation{Department of Physics, Indian Institute of Technology Guwahati, Guwahati 781039, Assam, India
}

\date{\today}

\begin{abstract}

	We present the global solutions of low angular momentum, inviscid, advective accretion flow around Kerr-Taub-NUT (KTN) black hole in presence and absence of shock waves. These solutions are obtained by solving the governing equations that describe the relativistic accretion flow in KTN spacetime which is characterized by the Kerr parameter ($a_{\rm k}$) and NUT parameter ($n$). During accretion, rotating flow experiences centrifugal barrier that eventually triggers the discontinuous shock transition provided the relativistic shock conditions are satisfied. In reality, the viability of shocked accretion solution appears more generic over the shock free solution as the former possesses high entropy content at the inner edge of the disc. Due to shock compression, the post-shock flow (equivalently post-shock corona, hereafter PSC) becomes hot and dense, and therefore, can produce high energy radiations after reprocessing the soft photons from the pre-shock flow via inverse Comptonization. In general, PSC is characterized by the shock properties, namely shock location ($r_s$), compression ratio ($R$) and shock strength ($S$), and we examine their dependencies on the energy (${\cal E}$) and angular momentum ($\lambda$) of the flow as well as black hole parameters.	We identify the effective domain of the parameter space in $\lambda-{\cal E}$ plane for shock and observe that shock continues to form for wide range of flow parameters. We also find that $a_{\rm k}$ and $n$ act oppositely in determining the shock properties and shock parameter space. Finally, we calculate the disc luminosity ($L$) considering free-free emissions and observe that accretion flows containing shocks are more luminous compared to the shock free solutions.

\end{abstract}

\maketitle

\section{Introduction}

Accretion process around compact objects is considered to be a fundamental 
mechanism \cite{Frank-etal2002} to explain the characteristics of the emergent electromagnetic radiation from different astrophysical sources like
quasars \cite{Smith-1966, Martin-2000}, active galactic nuclei \cite{Peterson-Bradley1997, Fabian-1999} and black hole X-ray binaries \cite{shakura-etal1973}. 
The effect of strong gravity resulted due to the 
central objects plays the crucial role in regulating such processes, and towards this, black hole assumes an esteemed position out of different possibilities of those central objects. In reality, the presence of horizon renders the black holes a unique place, where accretion phenomena arises naturally. And, in this realm, the relativistic hydrodynamics is the established framework which has been studied quite extensively in the literature \cite{Zanotti-2011}. Apart from understanding the underlying physical mechanism behind those astrophysical observations, accretion phenomena has recently taken an interesting direction in probing more generalized gravitational background \cite{Gyulchev-etal2020,Gelles-etal2021,Zeng-etal2022,Mirzaev-etal2022,Lin-etal2022,Liu-etal2022} due to recent observation of black hole shadow by the Event-Horizon Telescope(EHT) \cite{Akiyama-etal2019a, Akiyama-etal2019}. Due to the advent of state-of-the-art high precession observations in those front, it is timely to probe the different gravitational theory by means of the accretion flow dynamics. Meanwhile, numerous attempts along this line were carried out adopting different gravitational theories, such as higher dimensional braneworld gravity\cite{Pun-etal2008, Heydari-2010}, Chern-Simons modified gravity \cite{Harko-etal2010}, Ho{\v{r}}ava gravity \cite{lu-2009}, more exotic boson stars \cite{TORRES-2002, Guzman-2006}, wormholes \cite{Harko-etal2009a}, gravastars \cite{Harko-etal2009}, quark stars \cite{Kovacs-2009}. However, all these works were performed considering incomplete description of accretion flow, more specifically, taking into the account of the particle dynamics. Very recently, a full general relativistic hydrodynamic treatment is reported \cite{Dihingia-2020} for a special class of background called Kerr-Taub-NUT (KTN) spacetime \cite{Taub-1951, Newman-etal1963, Demianski-Newman1966}, and a complete set of accretion solutions and their properties are discussed. 

KTN spacetime is a bigger class of stationary and axisymmetric vacuum solution of the Einstein field equations ($R_{\mu \nu} -\frac{1}{2} g_{\mu \nu}R =0$, where $R_{\mu \nu}$ is Ricci tensor and $R$ is Ricci scalar) \cite{Misner-1963,Misner-Taub1969,Bonnor-1969,Miller-1973} with three independent parameters, namely mass $M_{\textrm{BH}}$, spin $a_{k}$, and NUT charge $n$. Further, the defining feature of such spacetime is that depending upon the value of the NUT parameter, spacetime contains either naked singularity or horizon with $a_k > 1$ as opposed to the usual Kerr black holes. The historical development of the KTN spacetime and its physical interpretations are delineated in the works of \cite{Taub-1951,Newman-etal1963,Misner-1963, Demianski-Newman1966}. Note that the NUT parameter $n$ is inferred as a semi-infinite massless rotating rod, which is situated along the rotational axis of the black hole. Accordingly, KTN spacetime is not asymptotically flat globally, however, it can be regarded as locally flat at spatial infinity \cite{Stelea-2006, Virmani-2011}. Generally, $n$ is interpreted as gravitomagnetic monopole, which is the gravitational analog of
a magnetic monopole. The effect of such exotic monopole is already examined in explaining several astrophysical findings, 
namely the spectra of supernovae, quasars, or active galactic nuclei \cite{Lynden-Nouri1998,Kagramanova-2010}, gravitational lens \cite{Nouri-Lynden1997}, microlens \cite{Rahvar-2003}, and black hole shadow \cite{Abdujabbarov-etal2013}.
Being motivated with this, in this work, we extend our previous analysis \cite{Dihingia-2020}, and for the first time to the best of our knowledge, study an important class of accretion solutions containing shock waves.

The formation of shock is a ubiquitous phenomena in nature, and accordingly, it is common in astrophysical environments as well. In the hydrodynamical framework, it appears as a discontinuous transition in terms of physical flow variables, such as velocity, density, pressure, and temperature. For accretion flows around black hole, 
phenomena of shocks has been studied quite extensively both theoretically \cite[and references therein]{Fukue-1987,Chakrabarti-1989, Abramowicz-etal1990, Das-etal2001a,Chakrabarti-Das2004,Das-2007,Fukue-2019a,Fukue-2019b,Das-etal2021} and numerically \cite[and references therein]{Molteni-etal1996,Das-etal2014,Okuda-Das2015,Sukova-etal2017}. For convergent accretion flow, the post-shock flow becomes hot and dense due to the shock compression and becomes puffed up resulting an effective boundary layer surrounding the black hole. Because of the excess temperature, the same post-shock flow (equivalently post-shock corona, hereafter PSC) acts as the source of hot electrons that eventually reprocesses the soft photons from the pre-shock flow via inverse Componization to produce high energy radiations \cite{Chakrabarti-Titarchuk1995,Mandal-Chakrabarti2005}. Following this, the existence of PSC is conjectured in examining the  emission from the active galactic nuclei (AGNs) and Galactic black hole sources (GBHs) \cite{rees-1984, grove-1998,Smith-etal2001,melia-2001,Smith-etal2002}. Meanwhile, \cite{Chakrabarti-Mandal2006,Nandi-etal2012,Iyer-etal2015,Nandi-etal2018} reported that PSC is possibly responsible for the spectral state transition in Galactic X-ray binary sources. 
The excess thermal energy across the shock front can also deflects a part of the accreting matter to produce jets/outflow along the direction of the black hole rotation axis \cite[and references therein]{Chakrabarti-1999,Das-Chakrabarti2008,Aktar-etal2015,Aktar-etal2017}. 
Further, time dependent numerical simulation studies
reveal that PSC undergoes aperiodic modulation as a consequence of resonant oscillation \cite{Molteni-etal1996,Das-etal2014} and therefore, the time varying undulation of PSC is possibly responsible for yielding the quasi-periodic oscillation (QPO) commonly seen in the power density spectra of Galactic black hole sources \cite{Craig-1999, Kluzniak-2001, Aschenbach-2004,Nandi-etal2012,Sreehari-etal2019,Sreehari-etal2020,Majumder-etal2022}.
Needless to mention that all these studies involving shock were carried out considering the standard black hole background. Hence, we wish to emphasize that it would be worthwhile to study the shock induced accretion flow dynamics considering more generalized KTN black hole spacetime.
  
As already indicated, our primary goal is to explore KTN spacetime as a special class of gravitational background, and study the effects of both Kerr ($a_{\rm k}$) and NUT ($n$) parameters on the shock properties, namely shock location ($r_s$), compression ratio ($R$), and shock strength ($S$). While doing so, we regard the full general relativistic hydrodynamic set up and consider relativistic equation of state (REoS) to take care the thermodynamical state of matter \cite{Chattopadhyay-Ryu2009}. To avoid complexity, we restrict our analysis in the adiabatic limit neglecting any dissipative processes, $i.e.$, viscosity, radiative cooling, magnetic fields. With this, we calculate the accretion solutions that are essential for shock formation and subsequently, obtain the complete set of shock-induced accretion solutions. As the global shocked solution has observational consequence, it is therefore timely to investigate the shock properties in terms of input parameters (flow parameters: energy ${\cal E}$ and angular momentum $\lambda$; KTN black hole parameters $a_{\rm k}$ and $n$). We identify the parameter space in $\lambda-{\cal E}$ plane for shock and find that shocked accretion solutions continue to exist for wide range of the input parameters. We further estimate the disc luminosities using the accretion solutions. Overall, in this paper, we discuss all these pertinent issues in depth and provide detail understanding of the shock-induced global accretion solutions around KTN black hole.

The paper is organized as follows. In \S II, we present the underlying assumptions and governing equations. In \S III, we discuss critical points. We present accretion solutions and shock properties in \S IV. In \S V, we obtain parameter space for shock and in \S VI, we calculate disc luminosity. Finally, in \S VII, we summarize our findings.

\section{Assumptions and governing equations}

We begin by considering a generic stationary axisymmetric back ground expressed as,
$$
\begin{aligned}
ds^2 = & g_{\mu\nu}dx^\mu dx^\nu\\
=& g_{tt}dt^2 + g_{rr}dr^2 + 2 g_{t\phi} dt d\phi + g_{\phi\phi} d\phi^2 +g_{\theta\theta} d\theta^2.
\end{aligned}
\eqno(1)
$$

In terms of Boyer-Lindquist coordinates \cite{Boyer-etal1967}, the components of the KTN metric are given by,
\begin{align}
g_{tt} &= (a_{\rm k}^{2}\sin^{2}\theta-\Delta)/\Sigma,\tag{1a}\\ \nonumber
g_{t\phi} &= -(A\Delta-a_{\rm k}B\sin^{2}\theta)/\Sigma,\tag{1b}\\ \nonumber
g_{rr} &= \Sigma/\Delta, \tag{1c}\\ \nonumber
g_{\theta\theta} &= \Sigma,\tag{1d}\\ \nonumber 
g_{\phi\phi} &= (B^{2}\sin^{2}\theta-A^{2}\Delta)/\Sigma. \tag{1e}
\end{align} 
Here, $\Delta= r^{2}-2r+a^{2}_{\rm k}-n^{2}$, $\Sigma = (a_{\rm k}\cos\theta+n)^{2}+r^{2}$, $A= a_{\rm k}\sin^{2}\theta-2n\cos\theta$, and $B= r^{2}+a^{2}_{\rm k}+n^{2}$, where $a_{\rm k}$ and $n$ are the Kerr parameter and NUT parameter, respectively. Setting $g^{rr}= 0$, we obtain the event horizon $(r_h)$ of the metric and is given by $r_{h}=1+\sqrt{1-a_{\rm k}^{2}+n^{2}}$. Clearly, based on the choice of $a_{\rm k}$ and $n$, the KTN spacetime yields either black hole for $(1-a^{2}_{\rm k} +n^{2} ) > 0$ or naked singularity when $(1- a^{2}_{\rm k} + n^{2} ) < 0$. Needless to mention that KTN spacetime reduces to the usual Kerr spacetime when NUT parameter vanishes ($n = 0$).

Further, we emphasize that the present hydrodynamic analysis is local and confined around the equatorial plane of the KTN space-time within the finite radius. Hence, the axial singularity mentioned in the Introduction does not affect the accretion flow dynamics.

In this paper, we use the sign convention as ($-, +, +, +$) and adopt the unit system $M_{\textrm{BH}} = G = c = 1$, where $M_{\textrm{BH}}$ is the mass of the black hole, $G$ is the universal gravitational constant, and $c$ is the speed of light. In this unit system, the length, angular momentum, and time are expressed in terms of $r_g~(=GM_{\textrm{BH}}/c^{2})$, $r_g c$ and $r_g/c$, respectively. 

The theoretical framework of our present study is mostly followed from our previous work \cite{Dihingia-2020}. Hence, without going into further technical details, we  straight away write down the dynamical equations and embark on calculating the shock induced global accretion solutions around KTN black holes. Towards this, we consider a geometrically thin advective accretion flow around a KTN spacetime in the steady state. Due to the axial symmetry of the background, we further assume that the accreting flow remains confined around the disc equatorial plane. At any point on the disc, the time-like four velocity of the flow is symbolized by $u^{\mu} \equiv (u^t,u^r,0,u^{\phi})$ and $u_\mu u^\mu = -1$. Following \cite{Dihingia-2020}, we obtain the relativistic Euler equation in the radial direction and is given by,
$$
v\gamma_{v}^{2}\frac{dv}{dr}+\frac{1}{h\rho}\frac{dp}{dr}+\frac{d\Phi^{\textrm{eff}}_{e}}{dr}=0.
\eqno(2)
$$
Here, $v$ denotes the radial three-velocity in the co-rotating frame and is defined as $v^{2}=\gamma^{2}_{\phi}v^{2}_{r}$, where $\gamma^{2}_{\phi}=1/(1-v^{2}_{\phi})$, $v^{2}_{\phi}=(u^{\phi}u_{\phi})/(-u^{t}u_{t})$, and $v^{2}_{r}=(u^{r}u_{r})/(-u^{t}u_{t})$. Further, $\gamma^{2}_{v} \left[ = 1/(1-v^{2})\right]$ is the radial Lorentz factor, $h\left[ =(e+p)/\rho\right]$ is the specific enthalpy, $e$ is the energy density, $p$ is the pressure, $\rho$ is the local mass density, and $\Phi^{\textrm{eff}}_{e}$ represents the effective potential \cite{Dihingia-2020} at the disc equatorial plane $(\theta=\pi/2)$ which is given by,

$$
\Phi^{\textrm{eff}}_{e}=1+\frac{1}{2}\ln\left[\frac{(n^{2}+r^{2})\Delta}{(a_{\rm k}^{2}-a_{\rm k}\lambda+r^{2}+n^{2})^{2}-(a_{\rm k}-\lambda)^{2}\Delta}\right],
\eqno(3)
$$
where $\lambda ~(= -u_{\phi}/u_{t})$ is the conserved specific angular momentum of the flow. For a stationary and axisymmetric KTN spacetime, there exists two commutating Killing vector fields, $l^{\mu}_t \equiv(\partial_{t},0,0,0)$ and $l^{\mu}_{\phi} \equiv(0,0,0,\partial_{\phi})$ corresponding to time translation and azimuthal rotation, respectively. The associated two conserved quantities along the direction of the motion are given by,

$$
hu_{\phi} = \mathcal{L}~({\rm constant});\quad  \quad -hu_{t} = \mathcal{E}~({\rm constant}),
\eqno(4)
$$ 
where $\mathcal{E}$ is the Bernoulli constant (equivalently specific energy). The mass conservation equation is given by,

$$
\dot{M}=-4\pi \rho r u^{r} H/\eta,
\eqno(5)
$$
where $\dot{M}$ refers the mass accretion rate which is treated as global constant. We express mass accretion rate in dimensionless form as ${\dot m}={\dot M}/{\dot M}_{\rm Edd}$, where ${\dot M}_{\rm Edd} ~(=1.44 \times 10^{18}$ gm s$^{-1}$) is the Eddington accretion rate. In equation (5), $\eta=r^{2}/(r^{2}+n^{2})$ and the half thickness ($H$) of the disc given by \cite{Riffert-Herold1995}, 

$$H=\sqrt{\frac{pr^{3}}{\rho {\cal F}} }; \quad
{\cal F}=\gamma_{\phi}^{2}\frac{(r^{2}+a_{\rm k}^{2})^{2}+2\Delta a_{\rm k}^{2}}{(r^{2}+a_{\rm k}^{2})^{2}-2\Delta a_{\rm k}^{2}}.\\$$
 
The entropy generation equation in the radial direction is obtained as,

$$
\left(\frac{e+p}{\rho}\right)\frac{d\rho}{dr}-\frac{de}{dr} = 0.\\
\eqno(6)
$$
In order to solve the governing equations (2), (4), (5) and (6), one requires a closure equation in the form of Equation of State (EoS), which describes the relation among the thermodynamical quantities like density ($\rho$), pressure ($p$) and internal energy ($e$).  During the course of accretion around the black holes, the accretion flow is expected to be in the thermally relativistic domain ($i.e.$, adiabatic index $\Gamma \rightarrow 4/3$) at the inner part of the disc, whereas it remains thermally non-relativistic ($i.e.$, $\Gamma \rightarrow 5/3$) at a distance far away from the black hole horizon \cite{Frank-etal2002}. Hence, the equation of state (EoS) with a fixed $\Gamma$ is inadequate to describe a thermally relativistic flow \cite{Taub-1948,Ryu-etal2006} and a relativistic EoS \cite{Chandrasekhar1939,Synge1957} needs to be employed, where $\Gamma$ is determined self-consistently based on the thermal properties of the flow. Keeping this in mind, we adopt an EoS for relativistic fluid  \cite{Chattopadhyay-Ryu2009} which is given by,
$$
e=\rho f \left(1+ \frac{m_p}{m_e}\right)^{-1}
$$
with
$$
f=\left[ 1+ \Theta \left( \frac{9\Theta +3}{3 \Theta +2}\right) \right] + \left[ \frac{m_p}{m_e} + \Theta \left( \frac{9\Theta m_e +3m_p}{3 \Theta m_e+ 2 m_p}\right) \right],
$$
where $\Theta~(=k_{\rm B}T/m_e c^2)$ is the dimensionless temperature, $m_e$ is the mass of electron, and $m_p$ is the mass of ion. According to the relativistic EoS, we express the speed of sound as $C_s = \sqrt{2\Gamma \Theta/(f+2\Theta)}$ \cite{Dihingia-etal2019}. Following \cite{Chattopadhyay-Ryu2009,Dihingia-2020} and using equation (5), we calculate the entropy accretion rate as,
$
\dot{\mathcal{M}}=\exp(k_1) \Theta^{3/2} \left(2 + 3 \Theta\right)^{3/4} \left(3 \Theta + \frac{2m_p}{m_e}\right)^{3/4} u^r r H/\eta,
$
where $k_1 = \left[ f - \left(1+m_p/m_e\right)\right]/2\Theta$.

We simplify equations (2), (4), (5), (6) and after some simple algebra, we obtain the wind equation as, 

$$
\frac{dv}{dr}=\frac{{\cal N}}{\cal{D}},
\eqno(7)
$$
where, ${\cal N}$ and ${\cal D}$ are respectively given by,

$$
{\cal N}=\frac{C_{s}^{2}}{\Gamma + 1}\left[\frac{1}{\Delta}\frac{d\Delta}{dr}-\frac{1}{\eta}\frac{d\eta}{dr}+\frac{3}{r}-\frac{1}{{\cal F}}\frac{d{\cal F}}{dr}\right]-\frac{d\Phi^{\textrm{eff}}_{e}}{dr},
\eqno(7a)
$$
and

$$
{\cal D}=\gamma_{v}^{2}\left[v-\frac{2C_{s}^{2}}{v(\Gamma + 1)}\right].
\eqno(7b)$$

Subsequently, using equations (2), (5), and (6) and employing the definition of sound speed, we calculate the differential form of the temperature as,

$$
\frac{d\Theta}{dr}=-\frac{\Theta}{2N+1}\left[\frac{\gamma_{v}^{2}}{2v}\frac{dv}{dr}+\frac{1}{\Delta}\frac{d\Delta}{dr}-\frac{1}{\eta}\frac{d\eta}{dr}+\frac{3}{r}-\frac{1}{{\cal F}}\frac{d{\cal F}}{dr}\right],
\eqno(8)
$$
where, $N \left[=(1/2) (df/d\Theta)\right]$ denotes the polytropic index of the flow \cite{Dihingia-etal2019} and it is related to the adiabatic index as $\Gamma = (1+N)/N$.

\section{Critical point and transonic solutions}

In an accretion process, gravity pulls matter towards the center of the gravitating object from the surrounding medium. In reality, inflowing matter starts its journey 
far away from the black hole with negligible radial velocity and enters into the black hole with velocity comparable to the speed of light just to satisfy the inner boundary conditions imposed by the event horizon. Because of this, the accretion around black holes bears an unique property that inevitably leads the flow to suffer smooth transition from subsonic to supersonic at the critical point before crossing the black hole horizon. Interestingly, depending on the flow parameters, namely energy and angular momentum, the flow may contain multiple critical points and thus is potentially viable to harbour shock waves as indicated in the Introduction. Before embarking on discussing shocks which is the main topic of the paper, let us briefly describe about the critical point, where both numerator and denominator of equation (7) simultaneously vanish as ${\cal N}={\cal D}=0$. Setting ${\cal D}=0$ in equation (7a), the radial flow velocity ($v_{c}$) at the critical point ($r_{c}$) is expressed as,

$$
v_{c}=\sqrt{\frac{2 }{\Gamma_{c} +1} }C_{sc},
\eqno(9)
$$
with the following expression for the sound speed,

$$
C^{2}_{sc}=\frac{\Gamma_{c} +1}{4}\left(\frac{d\Phi^{\textrm{eff}}_{e}}{dr}\right)_{c}
\left[\frac{1}{\Delta}\frac{d\Delta}{dr}+\frac{1}{\eta}\frac{d\eta}{dr}+\frac{3}{r}-\frac{1}{{\cal F}}\frac{d{\cal F}}{dr}\right]_c^{-1}
\eqno(10)
$$
In the equations (9-10), subscript `$c$' refers the flow variables measured at $r_c$. Since the flow remains smooth along the streamline including the critical points ($r_c$), the radial velocity gradient $(dv/dr)$ must be real and finite everywhere. Hence, we apply l$'$H\^{o}pital's rule to evaluate radial velocity gradient at $r_c$, and it generically possesses two types: $(dv/dr)_{c} < 0$ leads to the accretion solution while $(dv/dr)_{c} > 0$ yields winds. In the present work, we focus only on the accretion solutions keeping winds aside as they are of our particular interest. Depending upon the mathematical nature of $(dv/dr)_{c}$, critical points are classified in three categories. When $(dv/dr)_{c}$ assumes values which are real and of opposite sign, the critical point is called as saddle type. If $(dv/dr)_{c}$ are real and of same sign, it is called nodal type, and when $(dv/dr)_{c}$ are imaginary, it is called spiral type critical point. In the astrophysical context, saddle type critical points are of specially importance as the inflowing matter can only pass through them before entering into the black hole.

\begin{figure}
	\begin{center}
		\includegraphics[scale=0.25]{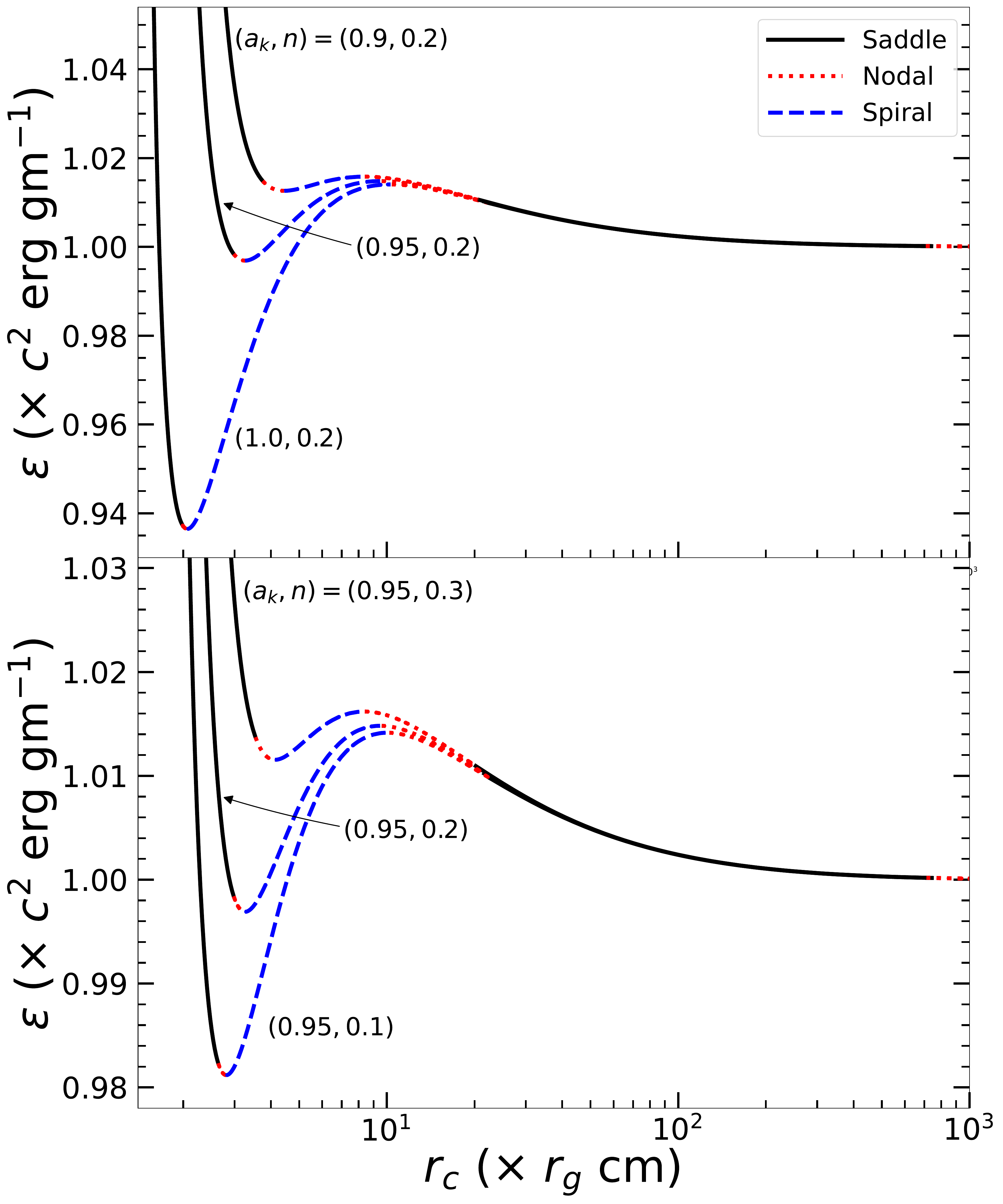}
	\end{center}
	\caption{Plot of energy ($\mathcal{E}$) as a function of critical points $(r_{c})$ for angular momentum $\lambda =2.04$. In upper panel, NUT parameter $n = 0.2$ and Kerr parameters are varied as $a_k = 0.9$, $0.95$ and $1.0$, while in the lower panel, $a_{\rm k}$ is kept fixed as $0.95$, and $n$ is varied as $0.1$, $0.2$ and $0.3$, respectively. In both panels, solid (black), dotted (red) and dashed (blue) curves represent the saddle, nodal and spiral type critical points and $a_{\rm k}$ and $n$ values are marked. See text for details. 
	}   
	\label{fig:1}
\end{figure}

In Fig. \ref{fig:1}, we show the variation of the flow energy (${\cal E}$) as a function of the critical points ($r_c$) for different values of $a_{\rm k}$ (upper panel) and $n$ (lower panel). In the figure, we choose $\lambda = 2.04$, and solid (black), dotted (red) and dashed (blue) curves represent the saddle, nodal and spiral type critical points, respectively. In the upper panel, we fix $n=0.2$ and vary $a_{\rm k}$ as $0.9$, $0.95$ and $1.0$ which are marked. We find that for a given $a_{\rm k}$, when $r_c$ is increased, the nature of the critical points is systematically altered following the sequence as saddle --- nodal --- spiral ---nodal--- saddle --- nodal. Moreover, we find that the range of multiple saddle type critical point locations required for shock transition increases as $a_{\rm k}$ is increased. We also find that there exists an upper limit of the location of the saddle type critical points for a relativistic flow accreting onto a KTN black hole. Similarly, in the lower panel, we choose $a_{\rm k}=0.95$, and vary NUT parameter as $n=0.1$, $0.2$ and $0.3$. Here, we observe that the role played by $n$ is characteristically opposite of $a_{\rm k}$ in deciding the properties of the critical points, however, the effect of one can not be completely negated by other due to the fact that KTN spacetime non-linearly depends on them.

In reality, depending on the input parameters ($\mathcal{E}, \lambda, a_{\rm k}, n$), accretion flow may contain maximum of three critical points outside the horizon \cite[]{Chakrabarti-1989,Das-etal2001a}. Among them, the inner critical points ($r_{\rm in}$) and the outer critical points ($r_{\rm out}$) generally form close to the horizon and far away from the horizon, whereas the middle critical points ($r_{\rm mid}$) form in between $r_{\rm in}$ and $r_{\rm out}$ \cite[]{Das-etal2001a}. Considering this, we present the change of critical point characteristics due to the variation of the input parameters in Table \ref{tab:table1}.
	
\begin{table}
\caption{Critical point locations and their natures obtained for various sets of input parameters (see Fig. \ref{fig:1}). See text for details.}
\begin{center}
\label{tab:table1}
		
\begin{tabular}{@{\hspace{0.1cm}} c @{\hspace{0.4cm}} c @{\hspace{0.4cm}} c @{\hspace{0.4cm}} c @{\hspace{0.4cm}} c @{\hspace{0.3cm}} l @{\hspace{0.4cm}} c @{\hspace{0.4cm}} r @{\hspace{0.0cm}} }\hline\hline
			
\multicolumn{3}{r}{Input Parameters} & & & \multicolumn{2}{r}{Critical Points} &\\	

			\cline{1-4}
			\cline{6-8}

$\lambda$ & $a_{\rm k}$  & $\mathcal{E}$ & $n$  &&   & \multicolumn{1}{l}{Location} & Type \\ 
{($cr_g$)} &  & ($c^2$) &  & &   & \multicolumn{1}{c}{($r_g$)} & \\ \hline

     &       &        &      && $r_{\rm in}$  & $3.336$ & Saddle \\
2.04 & 0.95  & 1.0180 & 0.3  && $r_{\rm mid}$ & --      & -- \\
     &       &        &      && $r_{\rm out}$ & --      & -- \\ \hline
			
     &       &        &      && $r_{\rm in}$  & $3.539$ & Saddle \\
2.04 & 0.95  & 1.0150 & 0.3  && $r_{\rm mid}$ & $5.972$ & Spiral \\
     &       &        &      && $r_{\rm out}$ & $11.915$& Nodal \\ \hline
			
     &       &        &      && $r_{\rm in}$  & $3.825$ & Nodal \\
2.04 & 0.95  & 1.0130 & 0.3  && $r_{\rm mid}$ & $4.792$ & Spiral \\
     &       &        &      && $r_{\rm out}$ & $15.667$& Nodal \\ \hline
			
     &       &        &      && $r_{\rm in}$  & --      & -- \\
2.04 & 0.95  & 1.0100 & 0.3  && $r_{\rm mid}$ & --      & -- \\
     &       &        &      && $r_{\rm out}$ & $22.763$& Saddle \\ \hline
			
     &       &        &      && $r_{\rm in}$  & --      & -- \\
2.04 & 0.95  & 1.0001 & 0.3  && $r_{\rm mid}$ & --      & -- \\
     &       &        &      && $r_{\rm out}$ & $1022.410$& Nodal \\ \hline
				
     &       &        &      && $r_{\rm in}$  & $2.741$  & Saddle \\
2.04 & 0.95  & 1.0050 & 0.2  && $r_{\rm mid}$ & $4.596$  & Spiral \\
     &       &        &      && $r_{\rm out}$ & $49.189$ & Saddle \\ \hline
			
     &       &        &      && $r_{\rm in}$  & $2.934$  & Saddle \\
2.04 & 0.95  & 0.9999 & 0.2  && $r_{\rm mid}$ & $3.832$  & Spiral \\
     &       &        &      && $r_{\rm out}$ & --       & -- \\ \hline
			
     &       &        &      && $r_{\rm in}$  & $2.996$  & Nodal\\
2.04 & 0.95  & 0.9990 & 0.2  && $r_{\rm mid}$ & $3.688$  & Spiral\\
     &       &        &      && $r_{\rm out}$ & --       & -- \\ \hline
			
		\end{tabular}
	\end{center}
\end{table}

\begin{figure}
	\begin{center}
		\includegraphics[width=\columnwidth]{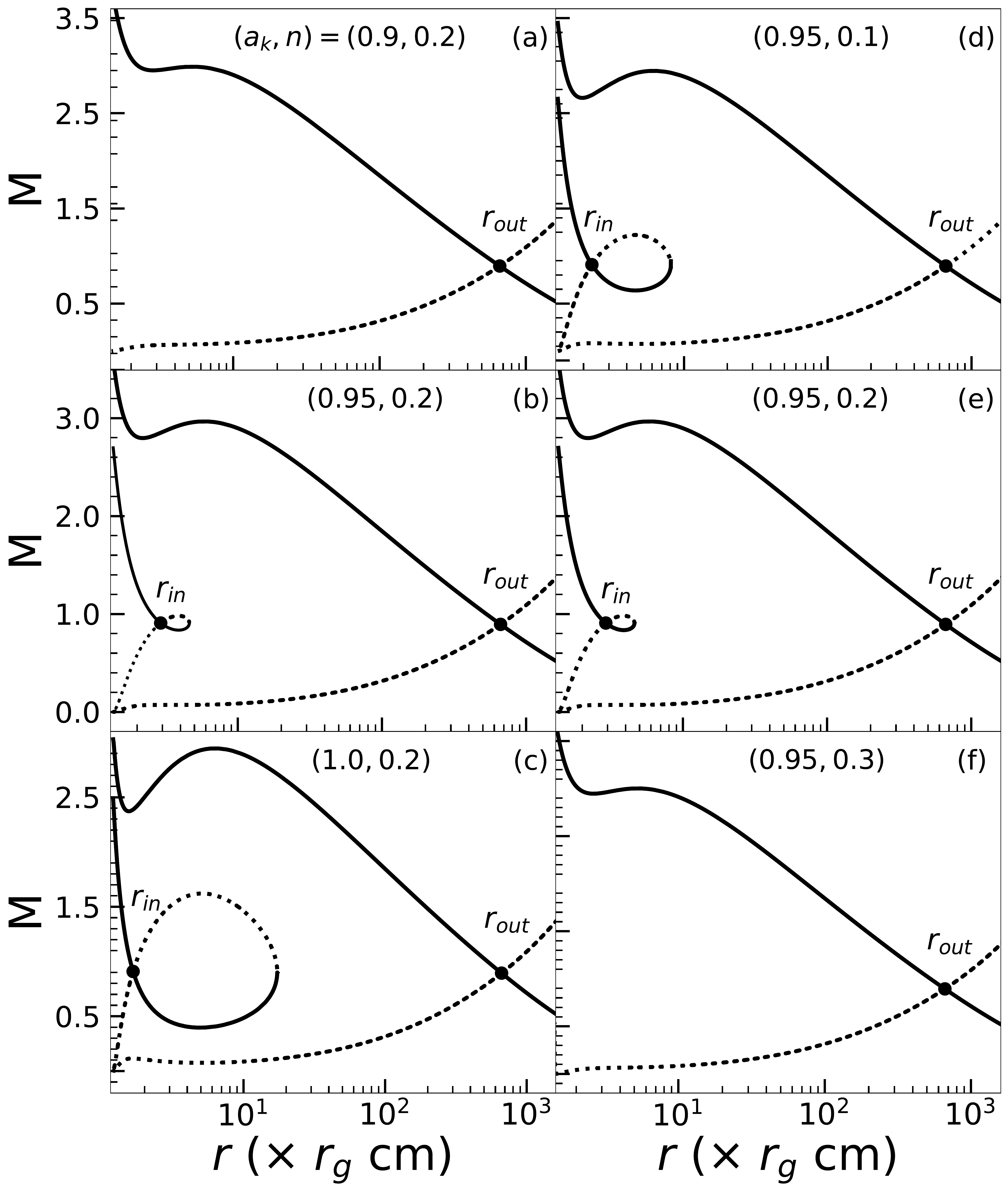}
	\end{center}
	
	\caption{Variation of Mach no $(M=v/C_{s})$ with the radial coordinate $(r)$ for energy $\mathcal{E}$ = 1.0002 and angular momentum $\lambda$ = 2.04. In the left panel, Kerr parameter is varied as $a_k = 0.9$ (top), $0.95$ (middle) and $1.0$ (bottom) keeping NUT parameter fixed as $n$ = 0.2. Similarly, in the right panels, Mach number is plotted for $n = 0.1$ (top), $0.2$ (middle) and $0.3$ (bottom), where $a_k$ = 0.95. The solid and dotted curves denote the accretion and wind solutions, respectively and filled circles refer the critical points.
	}   
	\label{fig:2}
\end{figure}

In order to understand the behavior of the flow containing saddle type critical points, in Fig. \ref{fig:2}, we provide representative plots of Mach number $(M=v/C_{s})$ as a function of radial co-ordinate ($r$) for $\mathcal{E} =1.0002$ and $\lambda = 2.04$. In the left panels, we investigate the effect of Kerr parameter ($a_k$) on the transonic solutions, whereas in the right panels, the role of NUT parameter ($n$) is examined. As mentioned earlier, we are interested to study the accretion solutions only, however, in each panel, we depict both accretion (solid curve) and its corresponding wind branch (doted) for the purpose of completeness. In Fig. \ref{fig:2}a, we choose $(a_{\rm k}, n)=(0.9,0.2)$ and flow contains only outer critical points at $r_{\rm out}=663.348$ and it successfully connects the black hole event horizon ($r_h$) with the outer edge of the disc ($r_{\rm edge} = 2000$). Here, filled circle denotes the critical point location. As Kerr parameter is increased keeping remaining parameters unchanged as $(a_{\rm k}, n)=(0.95,0.2)$, the inner critical point emerges (see Fig. \ref{fig:2}b) at $r_{\rm in}=2.917$. Interestingly, solution passing through $r_{\rm in}=2.917$ fails to connect the event horizon with $r_{\rm edge}$. With further increase of Kerr parameter as $(a_{\rm k}, n)=(1.0,0.2)$, inner critical point solution extends up to $\sim 1.660$ (see Fig. \ref{fig:2}c). The solution of such kind is potentially promising as it can join with another solution possessing $r_{\rm out}$ via discontinuous shock transition and ultimately provides a complete global accretion solutions connecting $r_h$ and $r_{\rm edge}$. We discuss the shock-induced global accretion solutions in the subsequent sections. In the right panels of Fig. \ref{fig:2}(d-f), examples of accretion solutions corresponding to  three different NUT parameters $n = 0.1 , 0.2, 0.3$ (marked in the figure) are presented, where $a_k = 0.95$ is kept fixed. We observe that NUT parameter ($n$) responds in the opposite manner on the nature of the flow portrait as the multiple saddle type critical points gradually disappear with the increase of $n$ for flows with fixed $a_{\rm k}$ values. Similar to the right panel, here also the solid and dotted curves denote accretion and wind solutions, respectively and the critical points are represented by filled circles.

\section{Accretion flow possessing shock}

As already mentioned that accreting matter around the black hole may contain multiple saddle type critical points depending on the flow parameters. In such cases, flow can not pass through both critical points simultaneously, unless a discontinuous transition of the flow variables happens at some radius in between the critical points. In reality, rotating accretion flow from the outer edge of the disc ($r_{\rm edge}$) starts accreting towards the black hole with negligible subsonic radial velocity and gradually gains its radial velocity as it proceeds towards the black hole. At $r_{\rm out}$, flow becomes transonic and continues to proceed further supersonically. Meanwhile, flow starts experiencing centrifugal repulsion that eventually slows down the inflowing matter causing its accumulation in the vicinity of the black hole and a centrifugal barrier is developed. The accumulation of matter does not continue indefinitely as the centrifugal barrier triggers the discontinuous transitions of the flow variables in the form of shock wave. Across the shock front, supersonic flow jumps into the subsonic branch and flow become compressed. Because of this, kinetic energy of pre-shock flow is converted into the thermal energy resulting the PSC. Usually, PSC acts as the source of hot and dense electrons where soft photons from the cooled pre-shock flow are intercepted and further reprocessed via inverse Comptonizationt to produce high energy radiations. After the shock transition, subsonic flow eventually picks up its radial velocity as it moves further inwards and again becomes supersonic after passing through $r_{\rm in}$ before crossing the black hole horizon. In order to execute such shock transition, accretion flow must satisfy the relativistic shock conditions \cite{Taub-1948}, which are given by,
$$
[\rho u^{r}]=0;\quad \left[(e+p)u^{t}u^{r}\right]=0; \quad\left[(e+p)u^{r}u^{r} + pg^{rr}\right]=0,
\eqno(11)
$$
where the quantities within the square bracket refer the differences of their values across the shock front. We utilize equation (11) to calculate the location of the shock transition ($r_s$).

\begin{figure}
	\begin{center}
		\includegraphics[width=0.95\columnwidth]{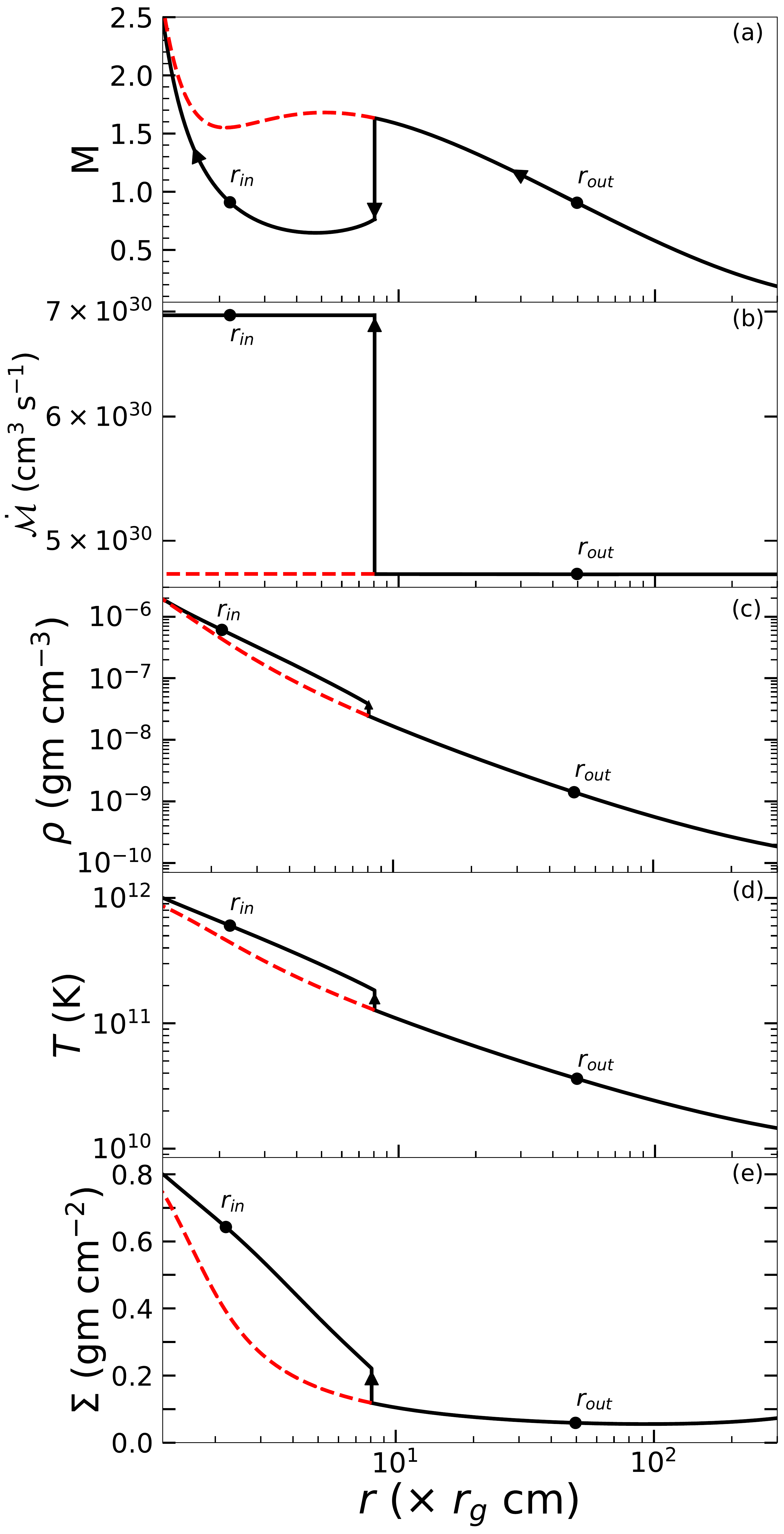}
		\caption{Variation of (a) Mach number $(M=v/C_{s})$, (b) entropy accretion rate $(\dot{\mathcal{M}})$, (c) density ($\rho$), (d) temperature $(T)$, and (e) surface density ($\Sigma$) as function of radial coordinate ($r$). Results are obtained for $({\cal E}, \lambda, a_{\rm k}, n)=(1.005, 2.1, 1.1,0.5)$, where solid (black) and dashed (red) curves represent shocked and shock free solutions. In each panel, filled circles denote the critical points ($r_{\rm in}$ and $r_{\rm out}$) and vertical arrow indicates the location of the shock transition. See text for details.
		}
		\label{fig:3}
	\end{center}
\end{figure}

\subsection{Typical Global accretion solutions with shock}

In Fig. \ref{fig:3}, we present a typical example of the shock-induced global accretion solution, where the variation of (a) Mach number ($M$), (b) entropy accretion rate $(\dot{\mathcal{M}})$, (c) density $(\rho)$, (d) temperature ($T$), and (e) surface density ($\Sigma$) are plotted as function of radial coordinate. Here, we choose the flow parameters as $(\mathcal{E}, \lambda) = (1.005, 2.1)$, and KTN black hole parameters as $(a_{k}, n) = (1.1, 0.5)$. In the figure, subsonic flow starts accreting from $r_{\rm edge} = 300$ and becomes supersonic after crossing $r_{\textrm{out}} = 49.65752$ (see Fig. \ref{fig:3}a). Subsequently, supersonic flow satisfies the relativistic shock conditions (equation 11) at $r_s = 8.0599$ and joins with the subsonic branch that eventually passes through the inner critical point at $r_{\textrm{in}} = 2.19679$ before falling in to the KTN black hole. In the figure, arrows indicate the overall direction of the flow motion towards the black hole where vertical arrow denotes the location of the shock transition. Indeed, after crossing $r_{\rm out}$, accretion flow has the possibility to enter into the black hole supersonically as depicted by the dashed curve (red). However, since entropy $(\dot{\mathcal{M}})$ associated with the post-shock flow is higher than the pre-shock flow (see Fig. \ref{fig:3}b), the second law of thermodynamics affirms that the shock solution is preferred over the shock-free solution \cite{Becker-Kazanas2001}. In Fig. \ref{fig:3}c, we demonstrate the density profile ($\rho$) corresponding to the shocked accretion solution presented in Fig. \ref{fig:3}a. We observe that $\rho$ increases immediately after the shock transition. This happens as radial velocity $v$ drops down to subsonic value at PSC and inward mass flux remains conserved across the shock front (see equation (5)). We further display the temperature profile of the accretion flow in Fig. \ref{fig:3}d. As expected, most of the kinetic energy of the pre-shock matter is converted into the thermal energy across the shock front that eventually causes the heating of PSC. Because of this, the rise of post-shock temperature profile is clearly seen. In Fig. \ref{fig:3}e, we show the profile of surface density ($\Sigma = \rho H$) \cite{Frank-etal2002}, where $\Sigma$ assumes higher values at PCS due to the density compression across the shock front. Such a distinct feature of PSC is naturally favourable to contain the swarm of hot electrons that can inverse Comptonize the soft photons coming from the outer region of the disc to produce high energy radiations commonly observed in Galactic black hole sources. Accordingly, the shocked accretion solutions emerge as a promising candidate in explaining the observational signatures of black holes \cite{Chakrabarti-Titarchuk1995,Mandal-Chakrabarti2005,Nandi-etal2012,Das-etal2021}.

\subsection{Shock properties}

In order to study the accretion disc structure around KTN black hole, one requires overall four parameters in the present formalism. Among them, two are conserved quantities associated with the flow, namely $\mathcal{E}$ and $\lambda$, and the remaining two are $a_k$ and $n$ that describe the KTN black hole. As already pointed out in section A, when shock conditions are satisfied, shock-induced global accretion solutions are favored over shock free solution. Therefore, in this section, we intend to examine the various shock properties, namely shock location ($r_s$), compression ratio ($R$), and shock strength ($S$).

\begin{figure}
	\begin{center}
		\includegraphics[width=\columnwidth]{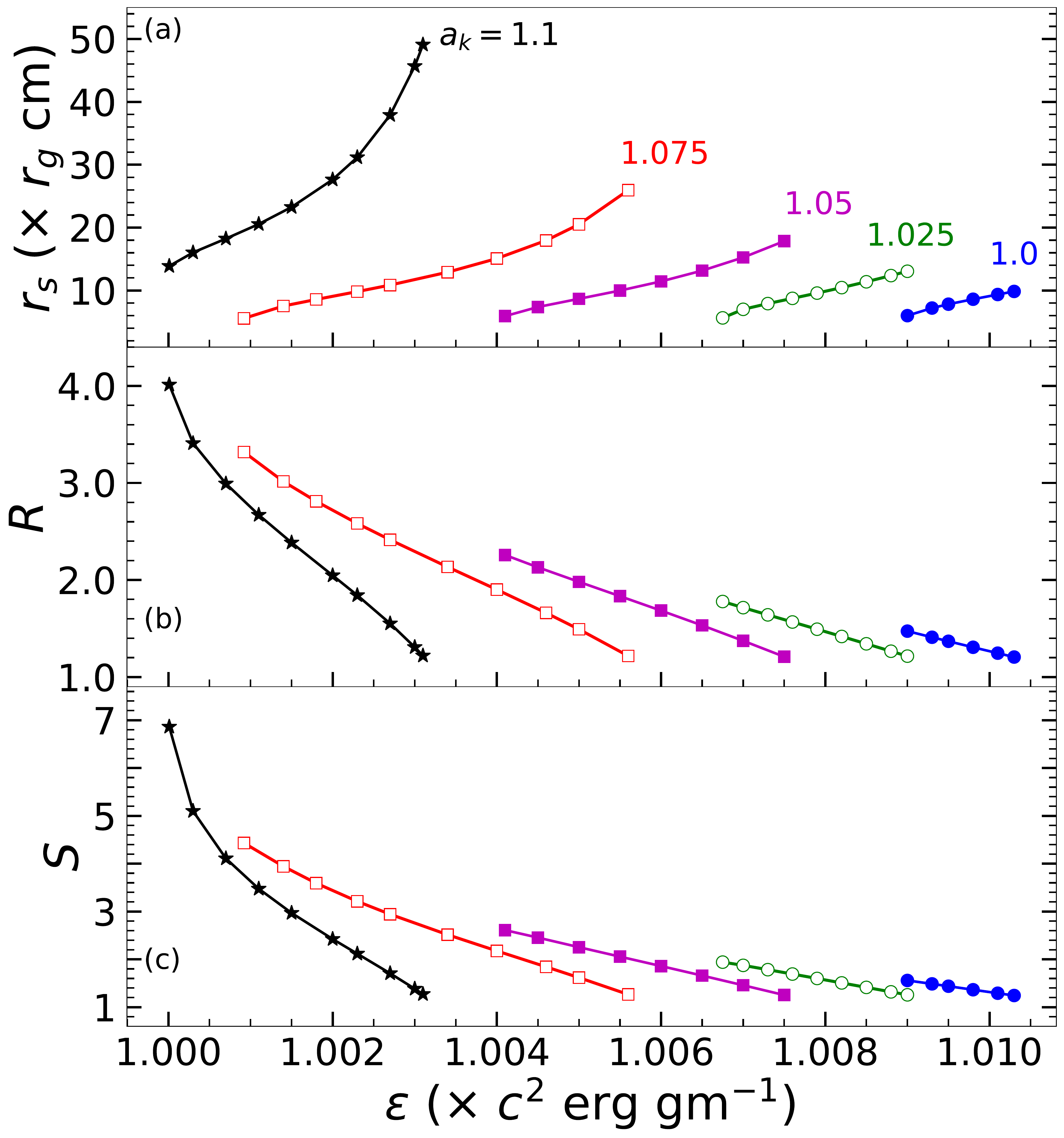}		
	\end{center}
	\caption{Variation of (a) shock location ($r_s$), (b) compression ratio ($R$), and (c) shock strength ($S$) as a function of specific energy ($\mathcal{E}$). Here, we vary Kerr parameter ($a_k$) from $1.00$ to $1.10$ with an interval $\Delta a_{\rm k}= 0.025$ from the right most curve to the left most curve. Here, we choose $\lambda = 2.2$ and $n = 0.5$. See text for details.
	}
	\label{rsRS_E_ak_Ln}
\end{figure}

Fig. \ref{rsRS_E_ak_Ln}a shows the variation of shock location ($r_s$) as a function of specific energy ($\mathcal{E}$) for different values of Kerr parameter ($a_{\rm k}$). Here, we choose $\lambda=2.2$ and $n=0.5$. The results plotted with filled circles (blue), open circles (green), filled squares (magenta), open squares (red), and asterisks (black) are for black hole spin starting from $a_{\rm k} = 1.0$ with increment $\Delta a_{\rm k} = 0.025$, respectively. We find that for a given $a_{\rm k}$, shock conditions are satisfied for a minimum energy (${\cal E}_{\rm min}$), and shock continues to form at larger radii as ${\cal E}$ is increased until it ceases to exist beyond ${\cal E}_{\rm max}$. This clearly indicates that for a given set of ($a_{\rm k}, \lambda,n$), there exist a range of energy ${\cal E}_{\rm min} < {\cal E} < {\cal E}_{\rm max}$ that yields global accretion solutions possessing standing shock. We also find that for a given ${\cal E}$, shock forms at smaller radii for black holes with smaller $a_{\rm k}$. Since the flow is compressed across the shock transition (see Fig. \ref{fig:3}) and the emergent radiations from the disc strongly depends on the density \cite{Chakrabarti-Titarchuk1995,Mandal-Chakrabarti2005}, it is useful to compute the compression ratio ($R$), which is defined as the ratio of the vertical averaged post-shock density ($\Sigma_{+}$) to the pre-shock density ($\Sigma_{-}$). In Fig. \ref{rsRS_E_ak_Ln}b, we present the variation of $R$ as function of ${\cal E}$ for the same set of flow parameters as in Fig. \ref{rsRS_E_ak_Ln}a. For a given set of ($a_{\rm k}, \lambda,n$), $R$ monotonically decreases with the increase of ${\cal E}$. This happens because the size of PSC increases with ${\cal E}$ that weakens the density compression across the shock front. A cut off is observed when shock disappears. We further emphasize that one would obtain both strong shock ($R\sim 4$) and weak shock ($R \rightarrow 1$) by suitably tuning the input parameters. We continue the study of shock properties by introducing another quantity called shock strength ($S$) which is defined as the ratio of the pre-shock Mach number ($M_{-}$) to the post-shock Mach number ($M_{+}$). In reality, $S$ determines the temperature jump across the shock front. In \ref{rsRS_E_ak_Ln}c, we depict the variation of $S$ as function of ${\cal E}$ for the same set of flow parameters as in Fig. \ref{rsRS_E_ak_Ln}a. We find that for a given set of ($a_{\rm k}, \lambda,n$), $S$ is stronger when ${\cal E}$ is small and it gradually becomes weak as ${\cal E}$ is increased till shock continues to exist. 

In reality, since PSC is the source of hot electrons by nature, it reprocesses the soft photons via inverse Comptonization to produce high energy radiations. In addition, $R$ and $S$ collectively regulate the efficiency of the inverse Comptonization process \cite{Chakrabarti-Titarchuk1995}. Therefore, the spectral properties of the central source are expected to be strongly dependent on these observables. Accordingly, by tuning the model parameters, namely ${\cal E}$, $\lambda$, $a_{\rm k}$ and $n$, it is possible to account the various spectral states commonly observed in black hole X-ray binary sources
\cite{Nandi-etal2012, Nandi-etal2018}.

\begin{figure}
	\begin{center}	
		\includegraphics[width=\columnwidth]{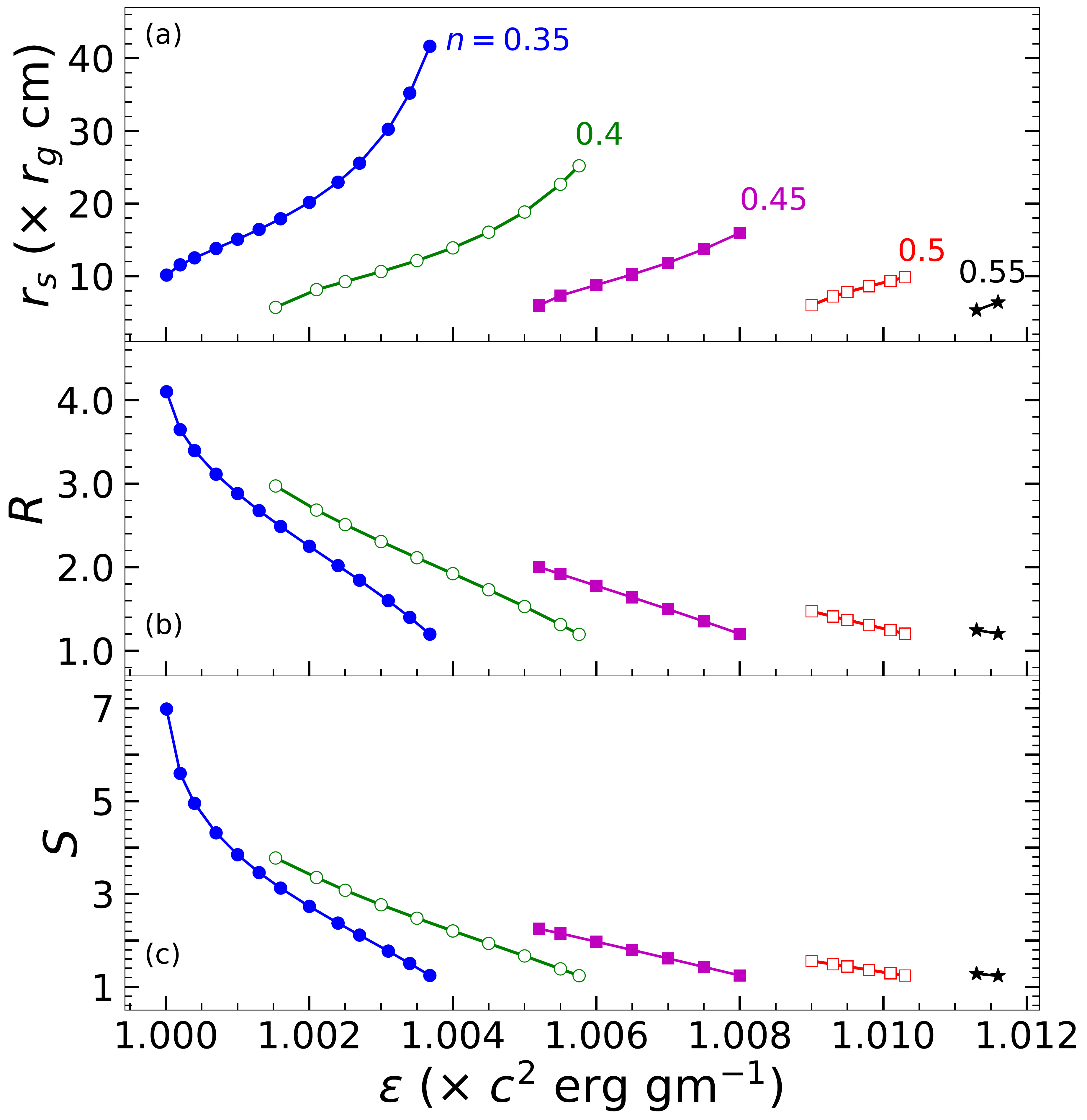}
	\end{center}
	\caption{Variation of (a) shock location ($r_s$), (b) compression ration ($R$), and (c) shock strength ($S$) as a function of specific energy ($\mathcal{E}$). Here, we vary NUT Parameter ($n$) from $0.35$ to $0.55$, with an interval $\Delta n= 0.05$ from the left most curve to the right most curve. We choose $\lambda = 2.2$ and $a_{\rm k} = 1.0$. See text for details.}
	
	\label{rsRS_E_n_Lak}
\end{figure}

In Fig. \ref{rsRS_E_n_Lak}, we present the variation of shock location ($r_s$), compression ratio ($R$), and shock strength ($S$) as a function of specific energy ($\mathcal{E}$). Here, we keep angular momentum and spin of the black hole fixed as $\lambda = 2.2$, and $a_{\rm k}=1.0$. The results plotted using filled circles (blue), open circles (green), filled squares (magenta), open squares (red), and asterisks (black)
are for $n = 0.35, 0.40, 0.45, 0.50$ and $0.55$, respectively. We observe that the role of $n$ in regulating the shock transition is entirely opposite to that of the black hole spin ($a_{\rm k}$) (see Fig. \ref{rsRS_E_ak_Ln}). Similar behaviours are seen for $R$ and $S$ as well. What is more is that for a given set of ($a_{\rm k}, \lambda$), shock-induced global accretion solutions are seen to exist for a wide range of NUT parameter ($n$).

\begin{figure}
	\begin{center}
		\includegraphics[width=\columnwidth]{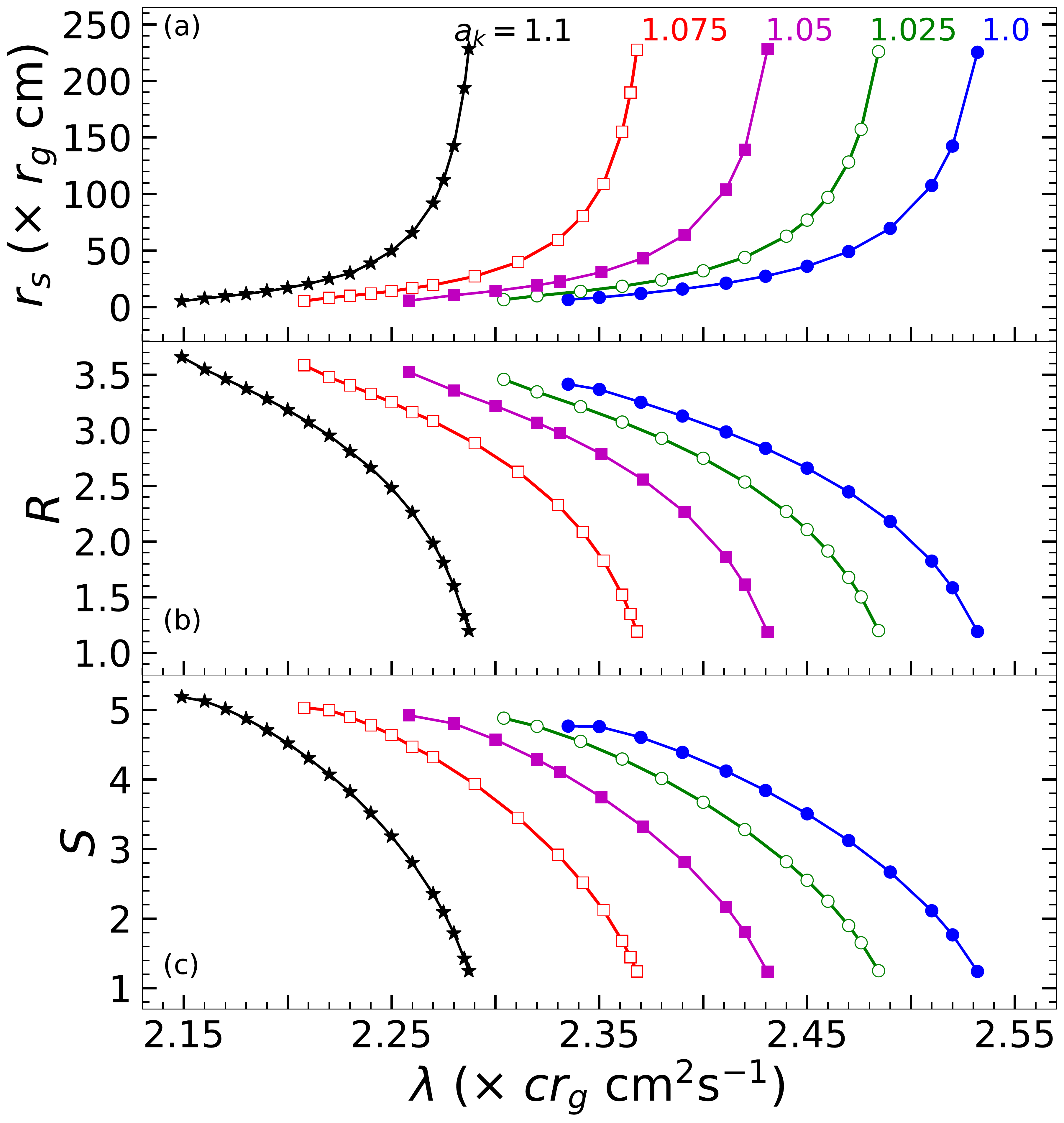}
	\end{center}
	\caption{Variation of (a) shock location ($r_s$), (b) compression ratio ($R$), and (c) shock strength ($S$) as a function of specific angular momentum ($\lambda$). Here, we vary Kerr parameter ($a_k$) from $1.0$ to $1.10$, with an interval $\Delta a_k = 0.025$ from the right most curve to the left most curve. The specific energy $\mathcal{E} = 1.0005$ and NUT parameter $n=0.5$. See text for details.}
	\label{rsRS_L_ak_En}	
\end{figure}

We continue the investigation of the shock properties and examine the role of angular momentum ($\lambda$) in regulating the shock dynamics. Towards this, we fix energy and NUT parameter as ${\cal E}=1.0005$ and $n=0.5$, and depict $r_s$, $R$ and $S$ as function of $\lambda$ for different $a_{\rm k}$ values in Fig \ref{rsRS_L_ak_En}a-c. Here, filled circles (blue), open circles (green), filled squares (magenta), open squares (red), and asterisks (black) denote the results obtained for $a_{\rm k}= 1.00$, $1.025$, $1.05$, $1.075$, and $1.10$, respectively. For a given $a_{\rm k}$, the shock forms further out for flows possessing higher $\lambda$. In reality, larger $\lambda$ causes enhanced centrifugal pressure that pushes the shock front outwards. In contrary, we find that shocks form for a particular range of $a_{\rm k}$, and for a given $\lambda$, the shock location decreases with the decrease of $a_{\rm k}$. Moreover, we observe that shock exists around black hole with higher $a_{\rm k}$, when the flow angular momentum is relatively lower and vice versa. This happens due to the spin-orbit coupling embedded within the space-time geometry where the marginally stable angular momentum is seen to decrease with the increase of $a_{\rm k}$ \cite{Das-Chakrabarti2008}. Consequently, the range of $\lambda$ for shock is also decreased with the increase of $a_{\rm k}$. Next, in Fig. \ref{rsRS_L_ak_En}b, we present the variation of compression ratio $R$ which are seen to decrease with $\lambda$ for black hole with identical spin value $a_{\rm k}$. In Fig. \ref{rsRS_L_ak_En}c, we plot the variation of the shock strength $S$ with $\lambda$, where similar variation is observed as in Fig. \ref{rsRS_L_ak_En}b.

\begin{figure}
	\begin{center}
	\includegraphics[width=\columnwidth]{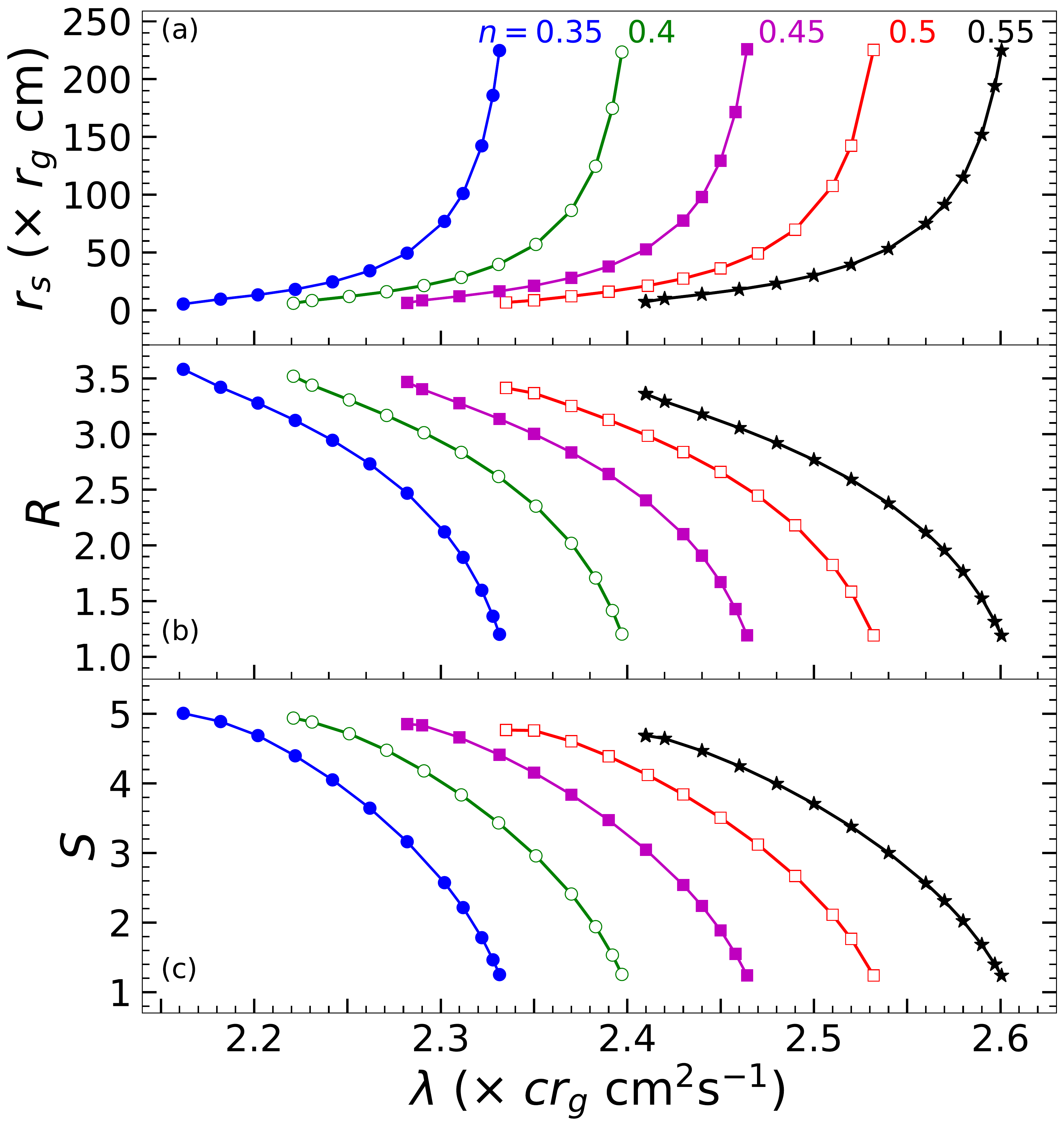}
	\end{center}	
	\caption{Variation of (a) shock location ($r_s$), (b) compression ration ($R$), and (c) shock strength ($S$) as a function of specific angular momentum ($\lambda$). Here, we vary NUT Parameter ($n$) from $0.35$ to $0.55$, with an interval $\Delta n = 0.05$ from the left most curve to the right most curve. We set $\mathcal{E}=1.0005$ and $a_k=1.0$. See text for details.}
	\label{rsRS_L_n_Eak}
	
\end{figure}

Fig. \ref{rsRS_L_n_Eak} illustrates the plot of (a) $r_s$, (b) $R$, and (c) $S$ with $\lambda$ for flows having different NUT parameter ($n$). Here, the remaining input parameters are chosen as ${\cal E} = 1.0005$ and $a_{\rm k} = 1.00$. In each panel, filled circles (blue), open circles (green), filled squares (magenta), open squares (red), and asterisks (black) denote the results corresponding to $n= 0.35, 0.40, 0.45, 0.50$ and $0.55$, respectively. Figure clearly indicates that shock solutions persist for a wide range of $n$. We find that $r_s$, $R$ and $S$ vary with $\lambda$ very similarly as in Fig. \ref{rsRS_L_ak_En}, however, as expected, the effect of $n$ on these observables is seen to be opposite of $a_{\rm k}$.

\begin{figure}[h]
	\begin{center}
		\includegraphics[width=\columnwidth]{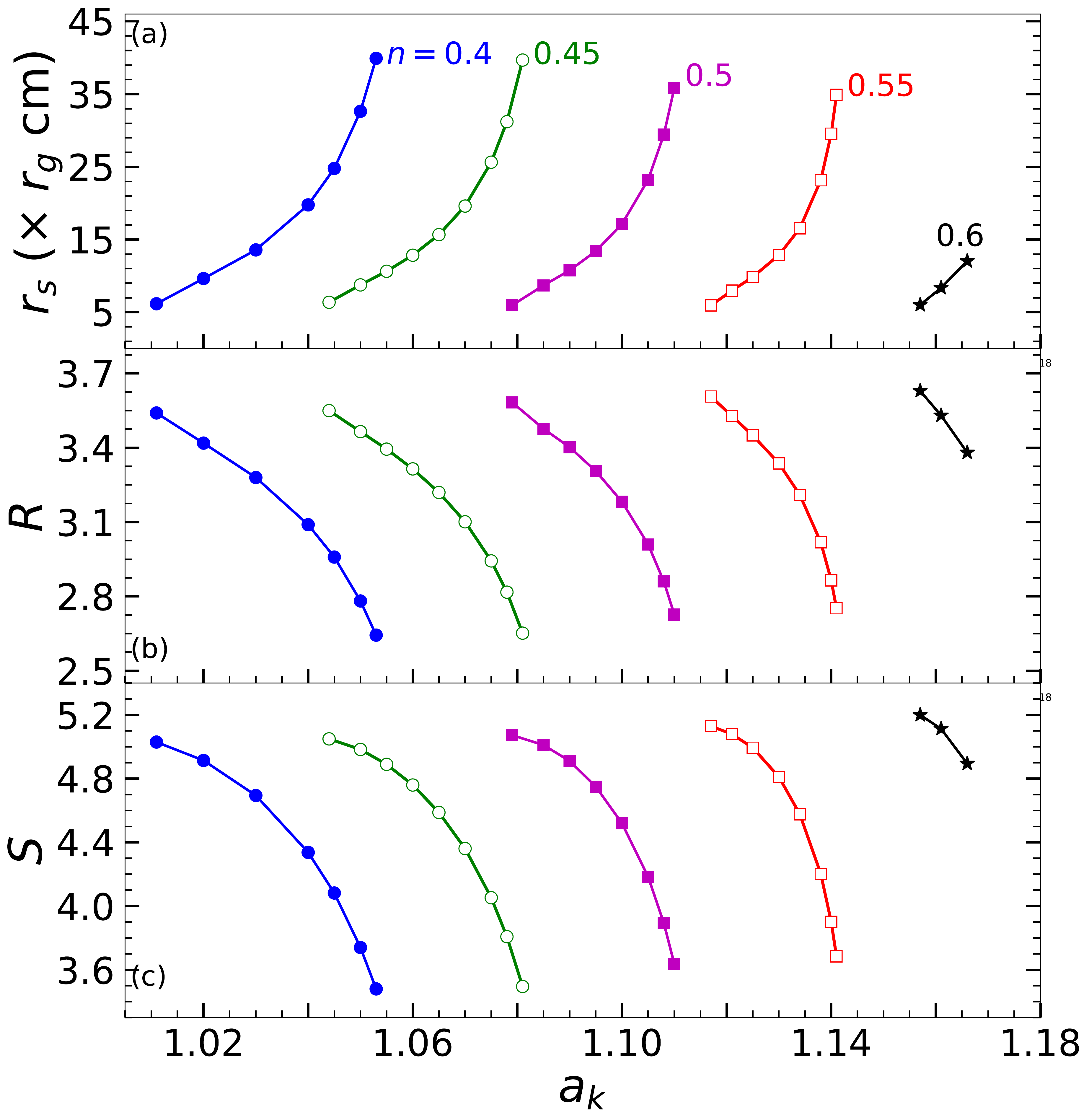}
	\end{center}
	\caption{Plot of (a) $r_s$, (b) $R$, and (c) $S$ as a function of $a_{k}$ for flows with $\mathcal{E} = 1.0005$ and $\lambda = 2.2$. In each panel, filled circles (blue), open circles (green), filled squares (magenta), open squares (red), and asterisks (black) joined using solid curves denote results obtained for $n = 0.4, 0.45, 0.5, 0.55$, and $0.6$, respectively. See text for details. 
	}
	\label{rs_ak_n_EL}
	
\end{figure}

Next, we examine how $r_s$, $R$ and $S$ vary with $a_{\rm k}$, when energy and angular momentum of the flow are kept fixed as ${\cal E} = 1.0005$ and $\lambda= 2.2$. The obtained results are shown in Fig. \ref{rs_ak_n_EL}, where each curve differs by $\Delta n = 0.05$ starting form the left most curve (filled circles, blue) with $n=0.4$. For a given set of $({\cal E}, \lambda, n)$, $r_s$ increases with $a_{\rm k}$, whereas both $R$ and $S$ decreases, as expected. We find that for a fixed $a_{\rm k}$, shocks form at smaller radii as $n$ is increased. It is evident from the figure that there exist a range $a_{\rm k}$ that admits shock solutions around KTN black hole and the said range explicitly depends on the other input parameters $({\cal E}, \lambda, n)$.

\begin{figure}[h]
	\begin{center}
		\includegraphics[width=\columnwidth]{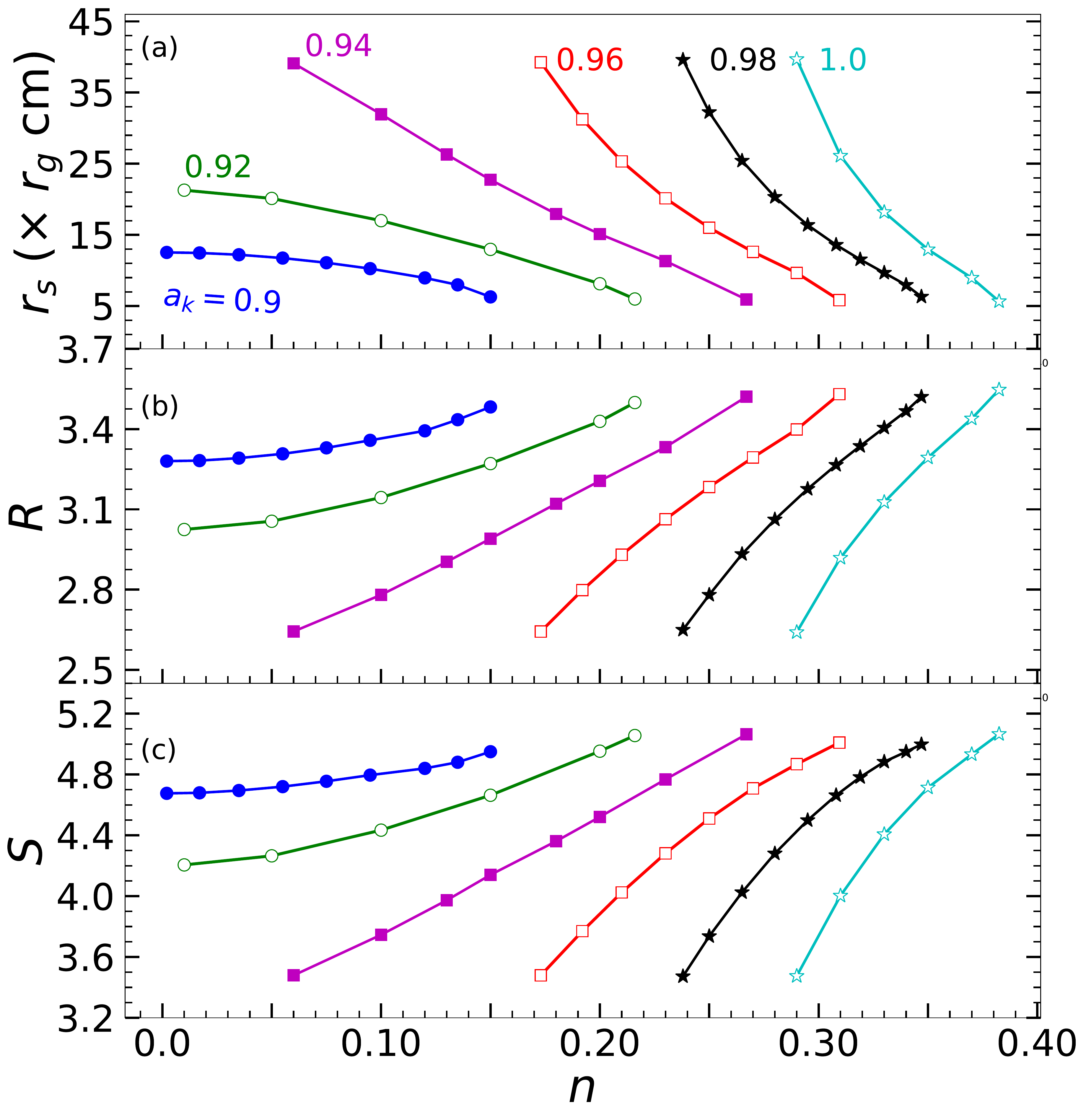}
	\end{center}

	\caption{Plot of (a) $r_s$, (b) $R$, and (c) $S$ as a function of $n$ for flows with $\mathcal{E} = 1.0005$ and $\lambda = 2.2$. In each panel, filled circles (blue), open circles (green), filled squares (magenta), open squares (red), filled asterisks (black), and open asterisks (cyan) joined using solid curves represent results for $a_{k} = 0.9, 0.92, 0.94, 0.96, 0.98$ and $1.0$, respectively. See text for details.
	}
	\label{rs_n_ak_EL}
\end{figure}

\begin{figure}
	\begin{center}
		\includegraphics[width=\columnwidth]{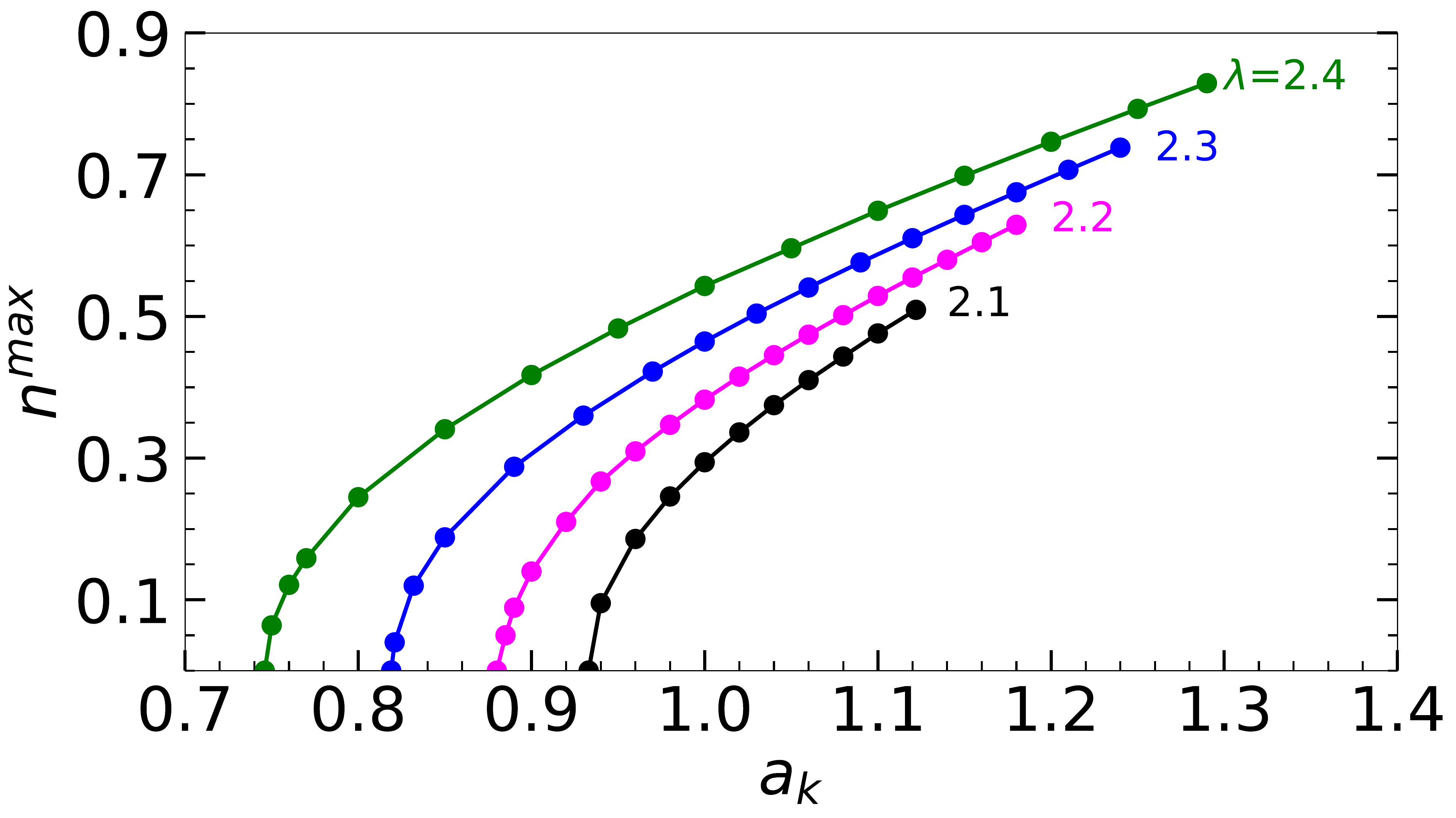}
			\end{center}
			
		\caption{Variation of the maximum value of NUT-parameter ($n^{\rm max}$) as a function of Kerr parameter ($a_k$) that admit shocks. Here, the flow energy is kept fixed as $\mathcal{E} = 1.0005$. Filled circles connected using solid curves in black, magenta, blue and green represent results for $\lambda = 2.1, 2.2, 2.3$, and $2.4$, respectively. See text for details.
		}
		\label{ak_nmax}
\end{figure}

In Fig. \ref{rs_n_ak_EL}, we show the variation of $r_s$, $R$ and $S$ with the NUT parameter ($n$) for rotating flows with ${\cal E}=1.0005$ and $\lambda=2.2$. Here, filled circles (blue), open circles (green), filled squares (magenta), open squares (red), filled asterisks (black), and open asterisks (cyan) represent the result for $a_{\rm k} = 0.9$, $0.92$, $0.94$, $0.96$, $0.98$ and $1.0$, respectively. We observe that for a fixed (${\cal E}, \lambda, a_{\rm k}$), $r_s$ in general decreases with $n$, and this eventually causes the increase of both $R$ and $S$ as well. It may be noted that accretion flow continues to experience shock transition around KTN black hole even for relatively high $n$ values. Evidently, when $a_{\rm k}$ is increased, the upper limit of NUT parameter ($n^{\rm max}$) for shock also increased. We quantify the correlation between $a_{\rm k}$ and $n^{\rm max}$ keeping energy of the flow fixed as ${\cal E}=1.0005$ and depict the obtained results in Fig. \ref{ak_nmax}. Filled circles joined using solid curves in black, magenta, blue and green denote the results corresponding to $\lambda = 2.1$, $2.2$, $2.3$, and $2.4$, respectively. It is apparent from the figure that for lower $a_{\rm k}$, $n^{\rm max}$ initially increases sharply and thereafter, it tends to sluggish at higher $a_{\rm k}$, in general.

\section{Parameter-space for Shock}

\begin{figure}
	\begin{center}
		\includegraphics[width=0.9\columnwidth]{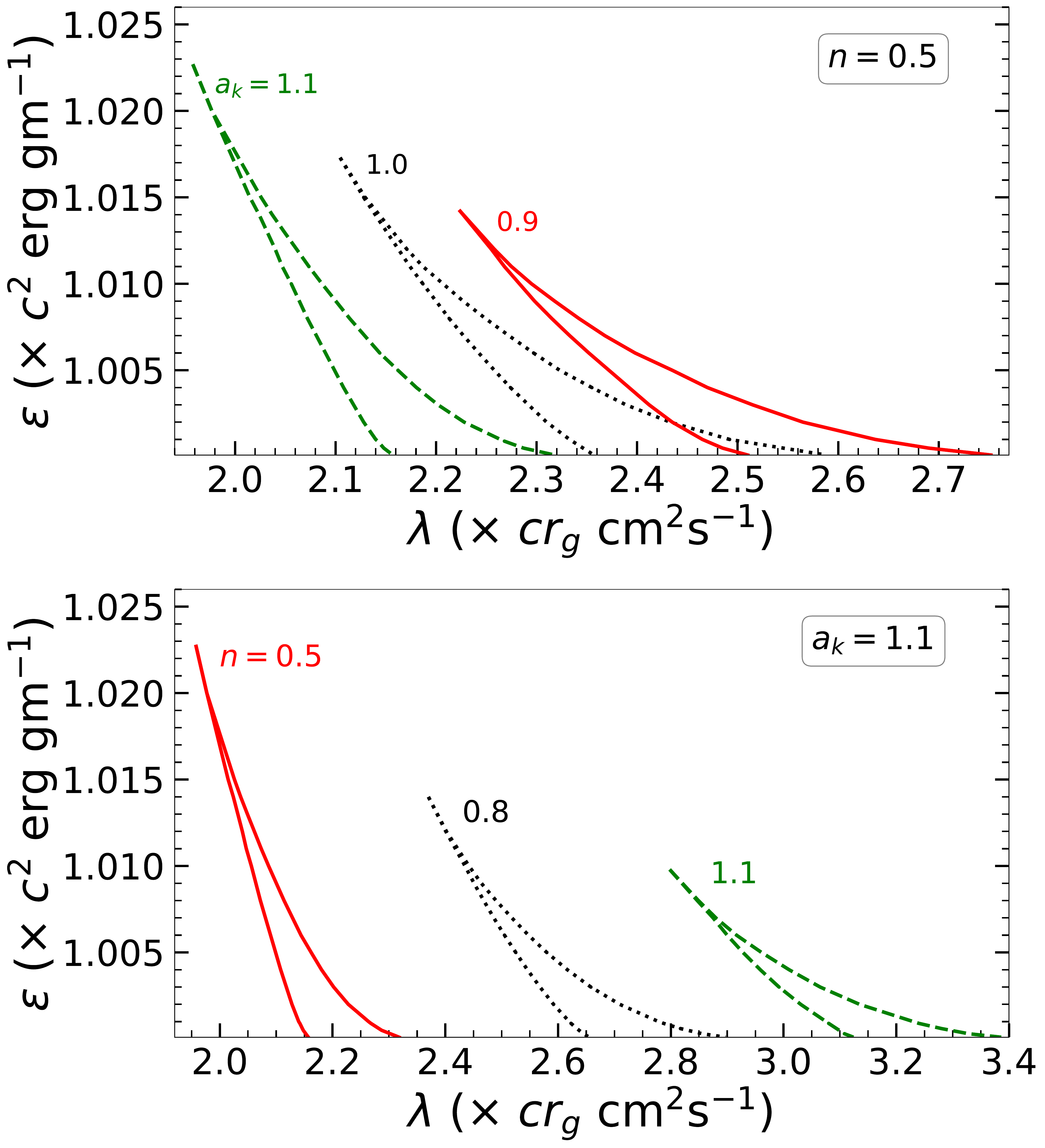}
	\end{center}	
	\caption{Modifications of the parameter space for shock in $\lambda-\mathcal{E}$ plane. In the upper panel, NUT parameter is set as $n = 0.5$ and the effective area bounded by red, black and green curves denote results for $a_{\rm k}=0.9, 1.0$, and $1.1$, respectively. In the bottom panel, Kerr parameter is chosen as $a_{\rm k}=1.1$ and the region separated using red, black and green boundaries are obtained for $n = 0.5, 0.8$, and $1.1$, respectively. See text for details.
	}
	\label{L_E_ak_n}
	
\end{figure}

In this section, we study the parameter space spanned by angular momentum ($\lambda$) and energy $(\mathcal{E})$ that admits shock-induced global accretion solutions around KTN black hole. In the upper panel of Fig. \ref{L_E_ak_n}, we depict the modification of the parameter space for different $a_{\rm k}$, where NUT parameter is kept fixed as $n=0.5$. The effective region bounded by the solid (red), dotted (black) and dashed (green) curves are for $a_{\rm k}=0.9$, $1.0$ and $1.1$, respectively. Clearly, lower $a_{\rm k}$ renders shock formation for relatively large $\lambda$ values. 
This findings are evident as the weakly rotating KTN black hole effectively diminishes the influence of the flow angular momentum due to the inherent spin-orbit coupling present within the space-time geometry describing the accreting system \cite{Das-Chakrabarti2008}. Similarly, in the lower panel of Fig. \ref{L_E_ak_n}, we identify the effective domain of the parameter space for different NUT parameters $n$ keeping the black hole spin fixed as $a_{\rm k}=1.1$. The regions enclosed by the solid (red), dotted (black) and dashed (green) curves are obtained for $n=0.5$, $0.8$ and $1.1$, respectively. We observe that the parameter space for shock gradually reduces and shifted towards the higher $\lambda$ and lower ${\cal E}$ domain as $n$ is increased. Overall, it is indeed evident that $a_{\rm k}$ induces effects opposite to $n$ in deciding the shock parameter space around KTN black holes.

\section{Radiative emissions in KTN space-time}

During the course of the accretion, the density ($\rho$) and temperature ($T$) of the infalling matter are increased as flow proceeds towards the black hole. This usually happens due to the overall geometrical compression experienced by a convergent flow. In addition, when the flow variables undergoes discontinuous transition, both density and temperature are enhanced further at the post-shock region due to shock compression. Because of these, the flow emits high energy radiations as the relevant radiative cooling mechanisms directly depend on $\rho$ and $T$ \cite{Das-2007,Dihingia-etal2020b}. As the accretion flow contains both species, namely electrons and ions, the free-free emission seems to be potentially viable. Hence, we consider bremsstrahlung radiation as an active radiative mechanism present inside the disc and calculate the total luminosity ($L$) as, 

$$
L=2 \int_{0}^{\infty}\int_{r_{h}}^{r_{\rm edge}} \int^{2\pi}_{0}(r + \eta^2/r)H\epsilon(\nu_e)d\nu_o dr d\phi.
\eqno(12)
$$
Here, $\epsilon(\nu)$ is the bremsstrahlung emission rate per unit volume per unit time per unit frequency \cite{Vietri-2008} and is given by,
$$
\epsilon(\nu)=\frac{32 \pi e^6}{3m_e c^3} \left(\frac{2 \pi}{3 k_B m_e T_e}\right)^{1/2} Z_i^2 n_e n_i g_{br} e^{-\left(h \nu / k_B T_e \right)},
\eqno(13)
$$
where $m_e$ and $e$ are mass and charge of the electron, $k_B$ is the Boltzmann constant, $h$ is the Planck’s constant, $\nu$ is the frequency, $Z_i$ is the charge of ion, and $g_{br}$ is the Gaunt factor \cite{Karzas-1961} assumed as unity in this work. We estimate the electron temperature as $T_e = \sqrt{(m_e/m_p)}T$ \cite{Chattopadhyay-Chakrabarti2002}, where $m_p$ is the ion mass and $T$ refers the flow temperature. The emitted frequency is obtained as $\nu_e = (1 + z)\nu_o$, where $\nu_o$ is the observed frequency and $z$
denotes the red-shift factor. Following the approach of \cite{Luminet-1979,Dihingia-2020}, we determine $z$ considering fixed inclination angle $i=\pi/4$ for the KTN black hole while calculating the disc luminosity (see also Appendix - A). In this work, we choose $r_{\rm edge}=1500$, and $M_{\rm BH}=10M_\odot$ all throughout unless stated otherwise.

\begin{figure}
	\begin{center}
		\includegraphics[width=\columnwidth]{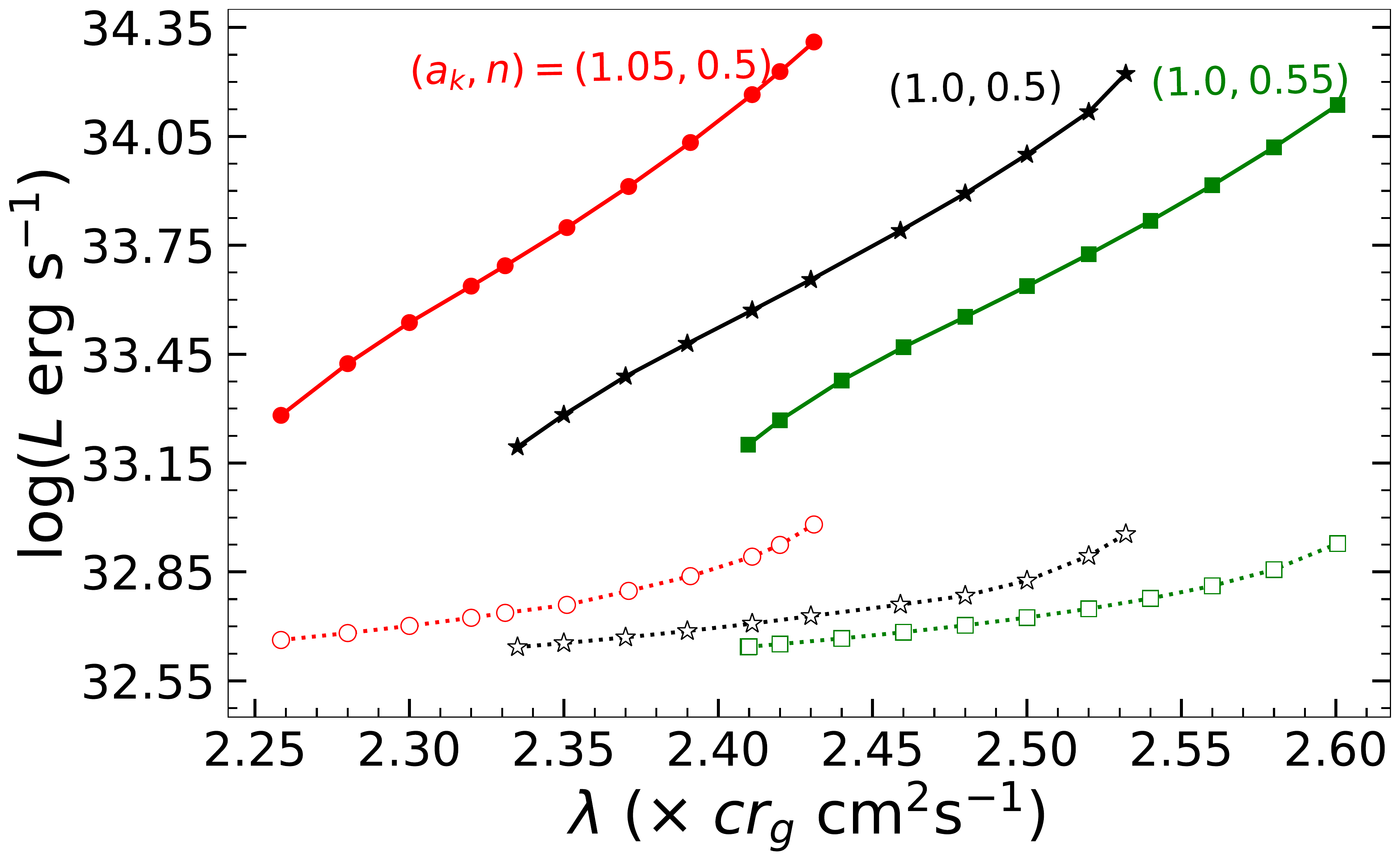}
	\end{center}
	
	\caption{Variation of disc luminosity ($L$) as a function of $\lambda$ for flows of energy ${\cal E} = 1.005$. Results depicted using filled circles, filled asterisks and filled squares are obtained for shocked accretion solutions, whereas open circles, open asterisks and open squares are for shock free solutions. Different points joined using red, black, and green solid curves are for ($a_{\rm k}, n$) = ($1.05,0.5$), ($1.00, 0.5$), and ($1.00, 0.55$), respectively. Here, we set ${\dot m} = 0.1$ and $M_{\rm BH}=10 M_\odot$. See text for details. 
	}
	\label{ak_Lu_L}
	
\end{figure}

In Fig. \ref{ak_Lu_L}, we present the variation of luminosity ($L$) with flow angular momentum $\lambda$ for various combination of ($a_{\rm k}, n$), which are marked. Here, we choose ${\cal E}=1.005$, and rate ${\dot m} = 0.1$. Filled points joined using solid curve
denote the results corresponding to the shocked accretion solutions, whereas open points joined by the dotted curve are for shock free solutions (see Fig. \ref{fig:3}). In the figure, it is evident that the disc luminosity for shocked accretion solutions always remains higher compared to the shock free solutions. Following this, we argue that the accretion solutions containing shock waves are potentially preferred in examining the energetics of the KTN black hole sources. We also find that for a given set of ($a_{\rm k}, n$), $L$ increases with $\lambda$ when flows harbor shocks, however, such increase is merely noticeable for flows having no shocks. In reality, for a given set of (${\cal E}, a_{\rm k}, n$), the shock radius (equivalently the size of the PSC) increases with $\lambda$ (see Fig. \ref{rsRS_L_ak_En} and \ref{rsRS_L_n_Eak}) and therefore, density and temperature profiles continue to remain higher within the PSC that eventually increase the disc luminosity. 

\section{conclusion}

In this work, we examine the structure of relativistic advective accretion flow around Kerr-Taub-NUT (KTN) black hole that possess shock waves. During the course of accretion, inflowing matter experiences the centrifugal repulsion that eventually triggered the shock transition. The dynamical shocked accretion solution overall includes the explicit dependence on energy (${\cal E}$) and  angular momentum ($\lambda$) of the flow as well as Kerr parameter ($a_{\rm k}$) and NUT parameter ($n$) of black hole. As part of the formalism, we adopt the relativistic equation of state (REoS) that satisfactorily describes the thermodynamical state of the accreting matter. The obtained results are summarized as follows.

\begin{itemize}

	\item We find that accretion flow around KTN black hole possesses either single or multiple critical points depending on the input parameters. We further observe that the nature of the flow solution alters as the spin parameter ($a_{\rm k}$) and NUT parameter ($n$) are varied (see Figs. \ref{fig:1}-\ref{fig:2}).
	
	\item When multiple critical points exist, accretion flow experiences discontinuous transition of the flow variables in the form of shock wave, provided relativistic shock conditions are satisfied (see Fig. \ref{fig:3}). We calculate the accretion solution containing shock waves and examine the shock properties, $i.e.$, shock location ($r_s$), compression ratio ($R$), and shock strength ($S$) in terms of the input parameters (Figs. \ref{rsRS_E_ak_Ln}-\ref{rs_n_ak_EL}).
		
	\item We notice that shock-induced global accretion solutions are not the isolated solutions, instead they exist for a wide range of the input parameters. We identify the effective domain of the parameter space in $\lambda-{\cal E}$ plane, that admits shocked accretion solutions. Further, we examine the modification of the parameter space due to the variation of ($a_{\rm k}, n$) and find that $a_{\rm k}$ and $n$ act oppositely in deciding the shock parameter space (see Fig. \ref{L_E_ak_n}). 
	
	\item We compute the maximum NUT parameter ($n^{\rm max}$) for shock as a function of $a_{\rm k}$, and find that shock solutions continue to exist for KTN black hole with $a_{\rm k}> 1$ as opposed to the Kerr black holes having $a_{\rm k} < 1$ (see Fig. \ref{ak_nmax}). 
	
	\item We calculate the disc luminosity ($L$) corresponding to the bremsstrahlung processes. We observe that $L$ is always higher for flows possessing shock waves compared to the shock free solutions (see Fig. \ref{L_E_ak_n}).
	
\end{itemize}

We further mention that the accretion solutions containing shocks can successfully render the spectral state transitions that are commonly observed in black hole X-ray binaries \cite{Chakrabarti-Titarchuk1995,Nandi-etal2018,Baby-etal2020}. Moreover, during the course of accretion, a part of the accreting matter can be deflected at PSC to produce bipolar jets/outflows due to the excess thermal gradient force present across the shock front \cite{Chakrabarti-1999,Das-etal2001b,Das-Chakrabarti2008,Aktar-etal2015,Aktar-etal2017}. What is more is that when PSC modulates, the quasi-periodic oscillations of the high energy photons in the spectral states is observed \cite{Chakrabarti-Manickam2000,Nandi-etal2001,Das-etal2021}. Considering all these, we presume that the shock-induced global accretion solutions seem to be potentially promising in explaining the observational findings of black hole sources \cite{Chakrabarti-Titarchuk1995,Iyer-etal2015,Sreehari-etal2019}.

Finally, we wish to emphasize that for the first time to the best of our knowledge, we obtain the shock-induced global accretion solutions around KTN black hole. Needless to mention that the present formalism is developed based on some simplifying assumptions. We neglect viscosity and radiative cooling mechanisms, namely synchrotron and Compton cooling processes, although they are expected to play viable role in regulating the disc dynamics. Moreover, we ignore the magnetic fields as well. Of course, the implementation of these physical processes are beyond the scope of the present paper and we intend to take them up for future works. In addition, we however argue that the basic conclusion of this work will not alter at least qualitatively upon involving all the physical processes mentioned above.

\section*{Acknowledgement}

Authors thank the anonymous reviewers for constructive comments and useful suggestions that help to improve the quality of the paper. This work was supported by the Science and Engineering Research Board (SERB) of India through grant MTR/2020/000331. GS acknowledges I. K. Dihingia for illuminating discussions and comments. Authors also thank the Department of Physics, IIT Guwahati, India for providing the infrastructural support to carry out this work.


\appendix

\section{Effect of general relativity on disc emission}

\begin{figure}
	\begin{center}
		\includegraphics[width=\columnwidth]{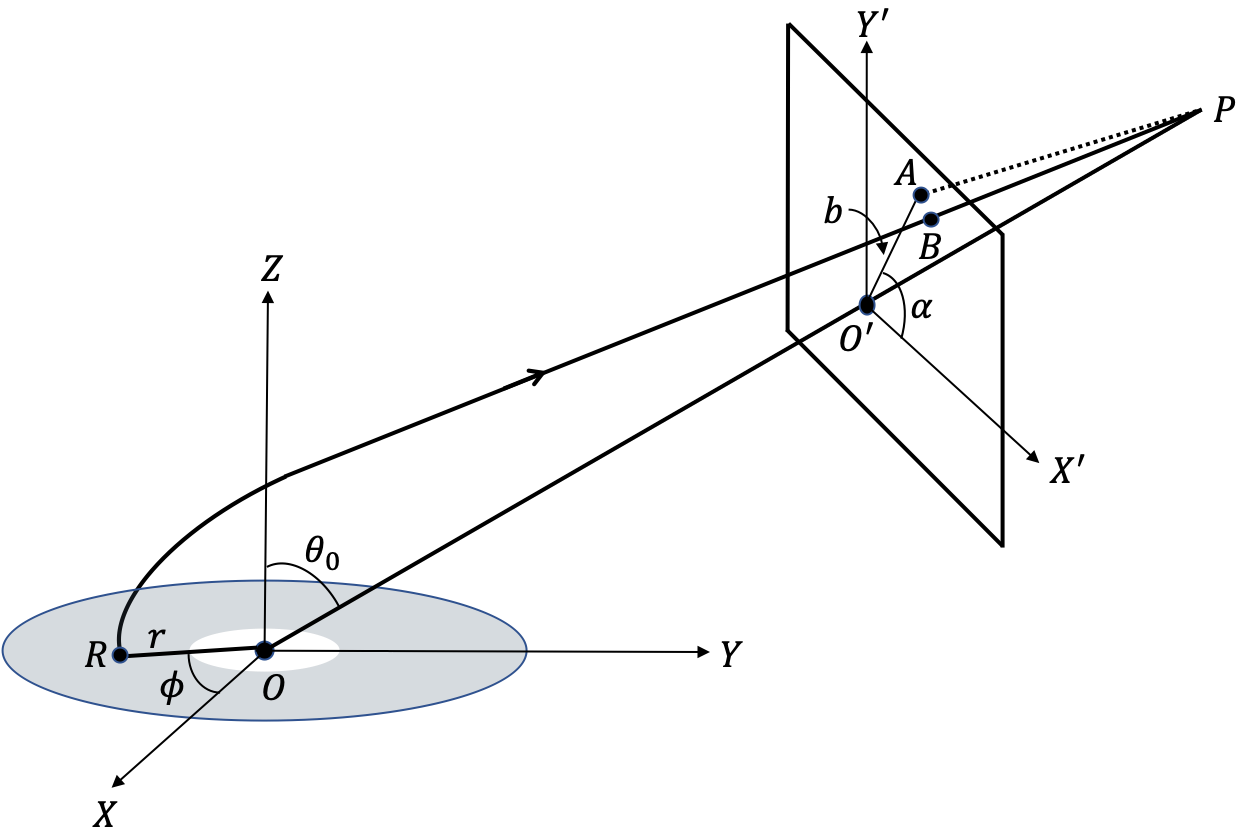}
	\end{center}
	\caption{Geometrical representation of a emitted photon from the accretion disc around a black hole. See text for details. 
	}   
	\label{fig:13}
\end{figure}

The frequency shift of an emitted photon due to the general relativistic effects is expressed in terms of redshift factor $z$ as,
\begin{equation}
\frac{\nu_{\rm o}}{\nu_{\rm e}}=\frac{E_{\rm o}}{E_{\rm e}} =\frac{(p_{\mu}u^{\mu})_{\rm o}}{(p_{\mu}u^{\mu})_{\rm e}}=\frac{1}{1+z},
\end{equation}
where, $\nu_{\rm o}$ is the observed frequency, $\nu_{\rm e}$ is the emitted frequency, $E_{\rm o}$ is the observed energy, $E_{\rm e}$ is the emitted energy, and $p_{\mu}$ is the $4$-momentum of the photon.

The energy of the emitted photon is given by,  
\begin{equation}
	E_{\rm e}= p_{t}u^{t} + p_{\phi}u^{\phi}=p_{t}u^{t}\left(1+\Omega\frac{p_{\phi}}{p_{t}}\right),
\end{equation} 
where, $\Omega~(=u^{\phi}/u^{t})$ is the angular velocity of the particle around the black hole, and $p_{\phi}/p_{t}$ refers the impact parameter of the photon around Z-axis (see Fig. \ref{fig:13}) which is given by,
\begin{equation}
	\frac{p_{\phi}}{p_{t}} = b\sin\alpha\sin\theta_{0}
\end{equation}

In Fig. \ref{fig:13}, the black hole is placed at the origin ($O$) of the XYZ-coordinate system. A photon is emitted from the point $R$ and being observed at the point $P$. $OP$ refers the distance between the observer and the black hole. The observer's screen is perpendicular to $OP$ and it is inclined with an angle $\theta_{0}$ to the Z-axis. On the observer's screen, $\rm (b,\alpha) $ are the coordinates of the intersection point of the tangential line to the photon trajectory. It is apparent that the effect of light bending generally diminishes provided the source of the emitted photon is located at far away distance from the observer. Considering this, we approximate $b \sin \alpha = r \sin \phi$ to calculate the redshift factor as,
\begin{equation}
1+z= u^{t}\big(1+\Omega r\sin\phi \sin\theta_{0} \big).
\end{equation}

We use equation (A4) to calculate the luminosity ($L$) corresponding to the bremsstrahlung radiation.

\bibliographystyle{unsrtnat}
\bibliography{ktn_shock_ref}

\end{document}